\documentclass[reprint,superscriptaddress,amsmath,amssymb,aps,floatfix,longbibliography]{revtex4-2}

\usepackage{graphicx}
\usepackage{color}
\usepackage{comment}
\usepackage{dcolumn}
\usepackage{upgreek}
\usepackage{gensymb}
\usepackage{float}
\usepackage{bm}
\usepackage{xcolor}
\usepackage{physics}
\usepackage{multirow}
\usepackage{array}
\usepackage{here}
\usepackage{tabularx}
\usepackage{threeparttable}
\usepackage{textcase}
\usepackage{hyperref}
\usepackage{xr}

\begin{document}
\title{Significant electron-magnon scattering in layered ferromagnet Cr$_2$Te$_3$}

\author{Yujun Wang}
\affiliation{Department of Physics, The University of Tokyo, Tokyo 113-0033, Japan}

\author{Shunzhen Wang}
\affiliation{Department of Physics, The University of Tokyo, Tokyo 113-0033, Japan}

\author{Masashi Kawaguchi}
\affiliation{Department of Physics, The University of Tokyo, Tokyo 113-0033, Japan}

\author{Jun Uzuhashi}
\affiliation{National Institute for Materials Science, Tsukuba 305-0047, Japan}

\author{Akhilesh Kumar Patel}
\affiliation{National Institute for Materials Science, Tsukuba 305-0047, Japan}

\author{Kenji Nawa}
\email[]{current affiliation: National Institute of Advanced Industrial Science and Technology, Tsukuba 305-8568, Japan}
\affiliation{Graduate School of Engineering, Mie University, Tsu 514-8507, Japan}
\affiliation{National Institute for Materials Science, Tsukuba 305-0047, Japan}

\author{Yuya Sakuraba}
\affiliation{National Institute for Materials Science, Tsukuba 305-0047, Japan}

\author{Tadakatsu Ohkubo}
\affiliation{National Institute for Materials Science, Tsukuba 305-0047, Japan}

\author{Hiroshi Kohno}
\affiliation{Department of Physics, Nagoya University, Nagoya 464-8602, Japan}

\author{Masamitsu Hayashi}
\email[]{hayashi@phys.s.u-tokyo.ac.jp}
\affiliation{Department of Physics, The University of Tokyo, Tokyo 113-0033, Japan}
\affiliation{Trans-scale quantum science institute, The University of Tokyo, Tokyo 113-0033, Japan}

\date{\today}

\begin{abstract}
A layered ferromagnet Cr$_2$Te$_3$ is attracting growing interest because of its unique electronic
and magnetic properties.
Studies have shown that it exhibits sizable anomalous Hall effect (AHE) that changes sign with temperature. 
The origin of the AHE and the sign change, however, remains elusive. 
Here we show experimentally that electron-magnon scattering significantly contributes to the AHE in Cr$_2$Te$_3$ through magnon induced skew scattering, and that the sign change is caused by the competition with the Berry-curvature or impurity-induced side-jump contribution. 
The electron-magnon skew scattering is expected to arise from the exchange interaction between the itinerant Te $p$-electrons and the localized Cr $d$-electrons modified by the strong spin-orbit coupling on Te. These results suggest that the magnon-induced skew scattering can dominate the AHE in layered ferromagnets with heavy elements. 
\end{abstract}
 
\maketitle

\clearpage
\section{Introduction}
The Cr-Te compound\cite{ipser1983jlesscommon} is a material system that has attracted significant interest recently owing to its unique structural, transport and magnetic properties. 
Many of the compounds form a layered structure and are stable down to a monolayer.
The majority of the compounds exhibit strong ferromagnetism with some exceptions (e.g. antiferromagnetism in CrTe$_3$\cite{mcguire2017prb} and Cr$_{1+\delta}$Te$_2$\cite{fujisawa2020prm}).
Studies have shown that ferromagnetism persists down to a monolayer\cite{zhang2021ncomm}, allowing studies on two-dimensional magnetism\cite{huang2017nature,gong2017nature,gibertini2019nnano}.
The Curie temperature typically lies in a range of 100 K to 200 K.
With proper growth conditions\cite{zhong2022nanores, li2019acsapplnanomater}, however, recent reports show the Curie temperature can be increased, exceeding room temperature under certain circumstances\cite{zhang2021ncomm, wen2020nanolett, chua2021advmater}.
Owing to the crystalline anisotropy, the magnetic easy axis often points along the film normal, with the perpendicular magnetic anisotropy energy larger than that of other layered ferromagnets\cite{coughlin2020acsnano, bian2021mrl}.

The transport properties of the compounds also show unique characteristics.
In particular, the compound exhibits a sizable anomalous Hall effect\cite{liu2018prb,jiang2020prb,huang2021acsnano}.
Studies have shown that the anomalous Hall resistance changes its sign as the temperature is varied\cite{li2019acsapplnanomater, sun2021aip, jiang2020prb, ou2022ncomm, cho2023nanoconv, chi2023ncomm, he2024advsci}.
The origin of the anomalous Hall effect as well as its sign change with temperature have been under scrutiny.
It has been reported that the anomalous Hall effect in the Cr-Te compounds is caused by the large Berry curvature of the bands near the Fermi level\cite{ou2022ncomm, chi2023ncomm, fujisawa2023advmater, he2024advsci, song2025advfuncmater}. As the Berry curvature induced anomalous Hall conductivity was found to be an odd function of energy, it causes a sign change in the anomalous Hall resistance as the temperature is varied due to population change of the occupied bands.
However, other studies\cite{liu2018prb, jiang2020prb, huang20212dmater} have shown that contribution from the skew scattering plays an essential role in the anomalous Hall effect, posing question on its origin.

Here we show that the unique characteristics of the anomalous Hall effect in Cr$_2$Te$_3$, one of the most stable compounds in the Cr-Te family, are defined by electron-magnon scattering.
Cr$_2$Te$_3$ has a layered structure in which layers of CrTe$_2$ are connected by intercalated Cr atoms.
We find the electron-magnon scattering significantly contributes to the longitudinal and anomalous Hall resistances.
The scaling relation between the longitudinal and anomalous Hall resistivities is used to identify the origin of the anomalous Hall effect.
We find two competing sources: magnon induced skew scattering and impurity induced side jump/Berry curvature effect.
Model calculations show that the former is caused by the exchange interaction between the itinerant Te $p$-electrons and the localized Cr $d$-electrons modified by the spin-orbit coupling.
We consider Te, which possess significant spin-orbit coupling, plays a critical role in setting the magnon-induced skew scattering.

\section{Experimental results}
\subsection{Structural and magnetic properties}
\begin{figure}[t]
	\centering
	\includegraphics[width=1.0\linewidth]{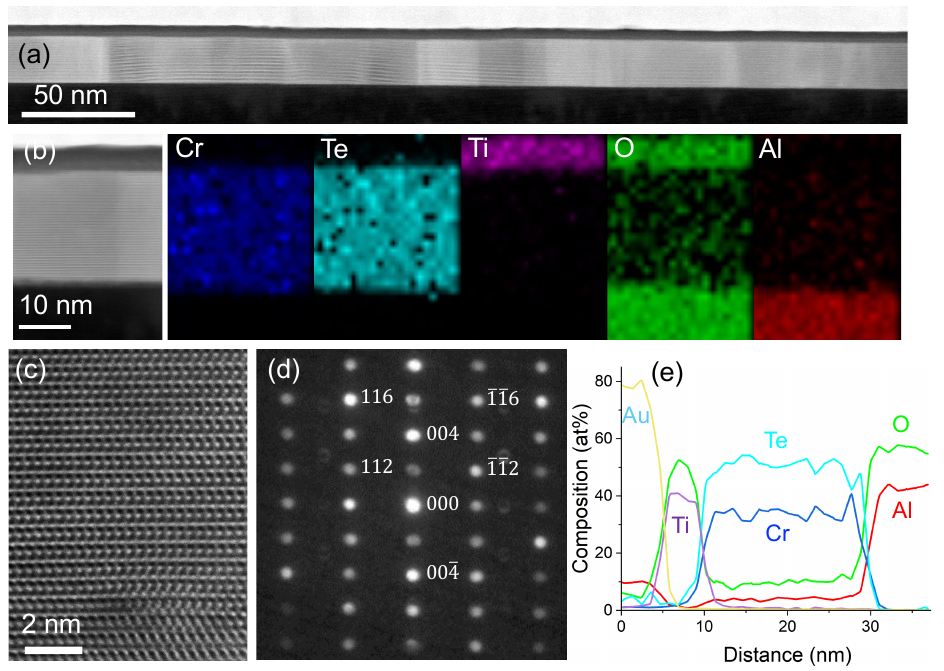}		
	\caption{\label{fig:structure} \textbf{Structural characterization.} (a) Cross-sectional high-angle annular dark-field scanning transmission electron microscopy (HAADF-STEM) image and (b) energy dispersive X-ray spectroscopy (EDS) maps of a 20 nm-thick Cr$_2$Te$_3$ film. (c) High magnification image of the Cr$_2$Te$_3$ film shown in (a). (d) Nanobeam electron diffraction pattern of the image shown in (c). Indices of Cr$_2$Te$_3$ are labeled. (e) Depth profile of the elements using EDS mapping. }
\end{figure}

Cr$_2$Te$_3$ films were grown on sapphire or MgO substrates using molecular beam epitaxy (MBE). A Ti or Te layer was used as a capping layer. 
See Sec.~\ref{sec:method:sample}
and Sec.~\ref{sec:supps:structure} 
for the details of sample preparation and characterization.
A cross-sectional high-angle annular dark-field scanning transmission electron microscopy (HAADF-STEM) image of a 20 nm-thick Cr$_2$Te$_3$ film is displayed in Fig.~\ref{fig:structure}(a). 
The bright and dark contrasts of the image represent grains with different crystal orientations within the film plane. The grain size is of the order of a few tens of nanometers. 
The high magnification image, Fig.~\ref{fig:structure}(c), and the corresponding nanobeam electron diffraction pattern, Fig.~\ref{fig:structure}(d), show highly textured film with growth along the Cr$_2$Te$_3$ (001) direction. 
Energy dispersive X-ray spectroscopy (EDS) maps of the elements are shown in Fig.~\ref{fig:structure}(b). The images show Cr and Te are uniformly distributed within the film. Profile of the film composition along the film normal is presented in Fig.~\ref{fig:structure}(e). From the profile, we determine the film composition is Cr:Te $\sim$ 2:3.
See Fig.~\ref{fig:rheed}
for the reflection high energy electron diffraction (RHEED) images and the X-ray diffraction (XRD) spectra of the films.

\begin{figure}[t]
	\centering
	\includegraphics[width=1.0\linewidth]{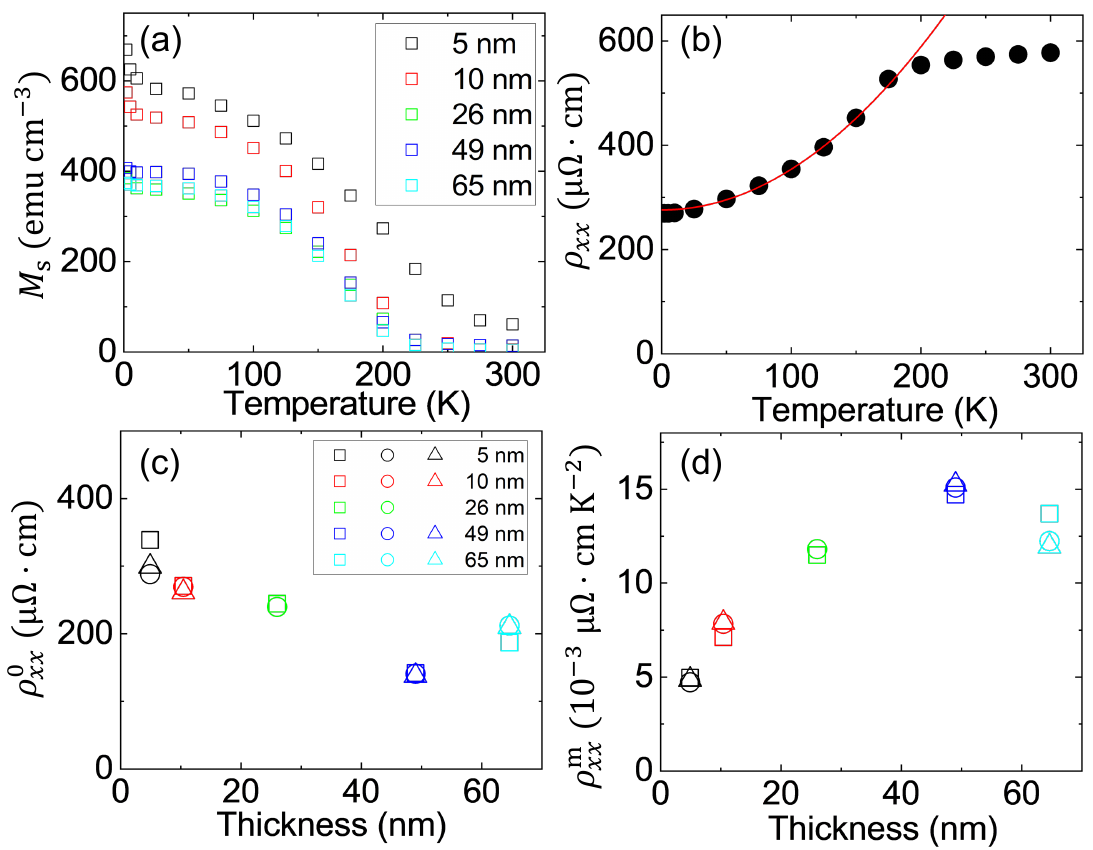}		
	\caption{\label{fig:rhoxx} \textbf{Saturation magnetization and longitudinal resistivity.} (a) Temperature $T$ dependence of the saturation magnetization $M_\mathrm{s}$ for films with different thicknesses. (b) Longitudinal resistivity $\rho_{xx}$ of a 10 nm-thick Cr$_2$Te$_3$ film plotted against $T$. The solid red line shows fit to the data using Eq.~(\ref{eq:rhoxx}) in the appropriate temperature range. 
    (c,d) Film thickness dependence of the impurity scattering coefficient $\rho_{xx}^0$ (c) and the electron-magnon scattering coefficient $\rho_{xx}^\mathrm{m}$ (d).
	For a given film thickness, results from a few devices are presented using different symbols. Definition of the symbols are the same for panels (c,d): see the legend shown in (c).}
\end{figure}

First, we study the magnetic properties of Cr$_2$Te$_3$.
Figure~\ref{fig:rhoxx}(a) shows the temperature dependence of the saturation magnetization $M_\mathrm{s}$ for films with different thicknesses.
The Curie temperature $T_\mathrm{C}$ is $\sim$175 K for all samples except for the 5 nm-thick film, which exhibits $T_\mathrm{C}$ of $\sim$215 K.
See  Sec.~\ref{sec:supps:trans}
for the details of how $T_\mathrm{C}$ is extracted. 
The value of $M_\mathrm{s}$ at 2 K for the thicker films is close to that predicted from first principles calculations, which is $\sim$465 emu/cm$^3$.
We find $M_\mathrm{s}$ is substantially larger for the thinner films (5 nm and 10 nm-thick).
Previous studies have reported that the magnetic moments of Cr$_2$Te$_3$ are canted from the film normal but the canting can be suppressed when the film thickness is reduced, thereby causing a difference in $M_\mathrm{s}$ against the film thickness\cite{bian2021mrl}.
Alternatively, it has been shown, using scanning tunnel microscopy, that Cr$_3$Te$_4$ forms at the beginning of the growth with MBE\cite{lasek2020acsnano}.
Since the saturation magnetization of Cr$_3$Te$_4$ is larger than that of Cr$_2$Te$_3$\cite{dijkstra1989jpcm, zhang2023advmater}, the magnetization can be larger for the thinner films given the larger weight of the Cr$_3$Te$_4$ phase. 
Note that the transport properties of Cr$_3$Te$_4$\cite{wang2022nanolett,yang20252dmater} are not significantly different from those of Cr$_2$Te$_3$.
In addition, $M_\mathrm{s}$ shows an upturn below $\sim$10 K for the thinner films. Such change in $M_\mathrm{s}$ at low temperature was reported previously\cite{li2019acsapplnanomater}.
Although the physical mechanism behind the upturn is unclear in Cr$_2$Te$_3$, previous studies for other systems (e.g. ultra-fine cobalt ferrite nanoparticles) suggested that it may originate from surface magnetic moments\cite{tung2003jap}.
Further study is required to clarify the origin of the thickness dependence of $M_\mathrm{s}$ and the upturn below $\sim$10 K.
In Fig.~\ref{fig:vsm},
we show a few exemplary magnetization hysteresis loops.
The loops show that the magnetic easy axis of the films points along the film normal, in agreement with previous studies\cite{huang20212dmater}.
For later use, we define $M_\mathrm{s}^0$ as $M_\mathrm{s}$ obtained at the lowest measurement temperature ($\sim$2 K).

\subsection{Longitudinal transport properties}
The transport properties are studied using the patterned Hall bars: 
see Sec.~\ref{sec:supps:trans} and Fig.~\ref{fig:measurement} for the details of the device structure.
The temperature dependence of the longitudinal resistivity $\rho_{xx}$ of a 10 nm-thick film is shown in  Fig.~\ref{fig:rhoxx}(b). 
We find the temperature dependence of $\rho_{xx}$ for $T < T_\mathrm{C}$ can be fitted with the following function: 
\begin{equation}
\begin{gathered}
\label{eq:rhoxx}
\rho_{xx} = \rho_{xx}^{0} + \rho_{xx}^\mathrm{m} T^2 \ \ \ (T<T_\mathrm{C}).
\end{gathered}
\end{equation}
The fitting result, shown by the red solid line in Fig.~\ref{fig:rhoxx}(b), is in good agreement with the experimental results. 
The change in $\rho_{xx}$ for $T > T_\mathrm{C}$ is almost linear and its slope is small: see also 
Fig.~\ref{fig:transport}.

The quadratic temperature dependence of $\rho_{xx}$ below $T_\mathrm{C}$ can be attributed to electron-magnon scattering\cite{mannari1959ptep,volkenshtein1973pssb}.
(As is often the case in metals\cite{volkenshtein1973pssb}, we neglect electron-electron scattering, which also scales with $T^2$.)
The temperature independent resistivity that is dominant at the lowest temperature is likely associated with impurity induced scattering.
We therefore assign $\rho_{xx}^0$ and $\rho_{xx}^\mathrm{m}$ as the impurity and electron-magnon scattering coefficients, respectively.
The film thickness dependences of $\rho_{xx}^0$ and $\rho_{xx}^\mathrm{m}$ are shown in Figs.~\ref{fig:rhoxx}(c) and \ref{fig:rhoxx}(d), respectively. 
$\rho_{xx}^0$ tends to decrease with film thickness, suggesting that the film quality improves for thicker films.
In contrast, $\rho_{xx}^\mathrm{m}$ increases with the film thickness until it saturates at $\sim$50 nm.
\begin{figure}[t]
	\centering
	\includegraphics[width=1.0\linewidth]{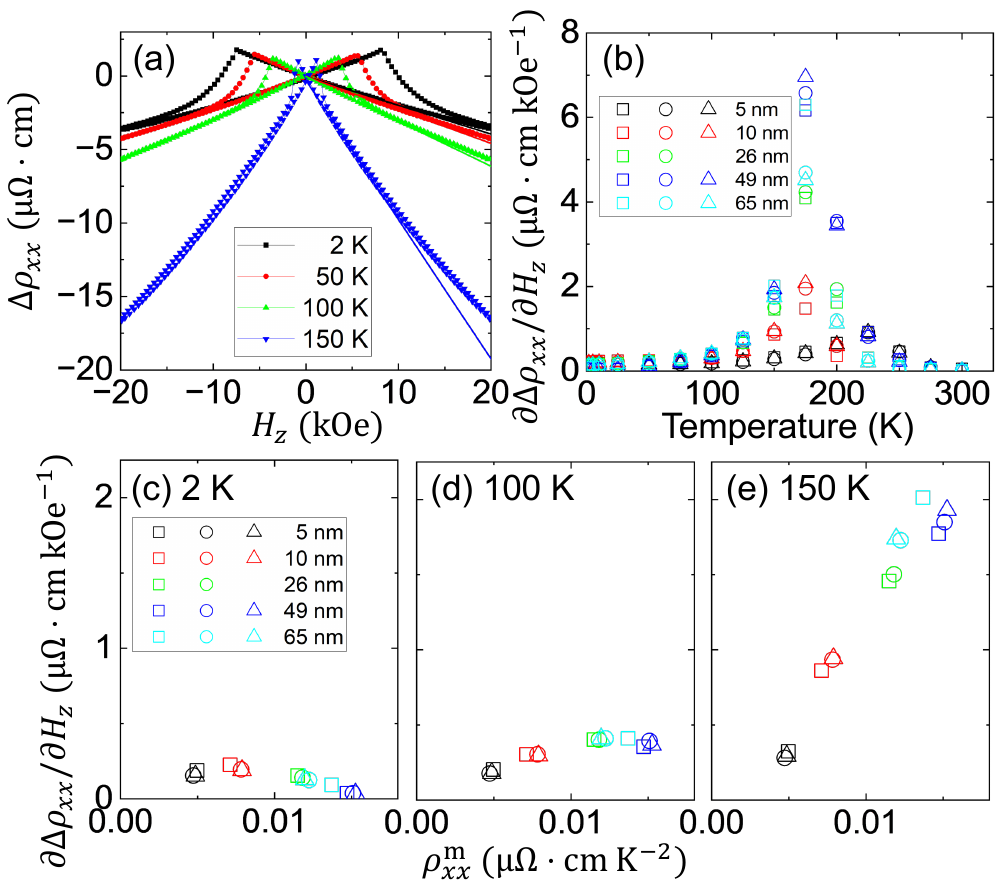}		
	\caption{\label{fig:mr} 
	\textbf{Longitudinal magnetoresistance.} (a) Magnetoresistance $\Delta \rho_{xx}$ as a function of out of plane magnetic field $H_z$ obtained for a 10 nm-thick film. Different symbols indicate different measurement temperatures. The solid lines show linear fits to the data when $H_z$ is in the range of 0 to 10 kOe. (b) Slope of fitted linear line, $\frac{\partial \Delta \rho_{xx}}{\partial H_z}$, plotted as a function of temperature. (c-e) Electron-magnon scattering coefficient $\rho_{xx}^\mathrm{m}$ dependence of $\frac{\partial \Delta \rho_{xx}}{\partial H_z}$ obtained at 2 K (c), 100 K (d) and 150 K (e). (b-e) For a given film thickness, results from a few devices are presented using different symbols. Definition of the symbols are the same for panels (c-e): see the legend shown in (c).}
\end{figure}

To show that electron-magnon scattering indeed contributes to the transport properties, the out of plane magnetic field ($H_z$) dependence of the longitudinal resistivity $\Delta \rho_{xx} \equiv \rho_{xx}(H) - \rho_{xx}(H=0)$ is studied.
Figure~\ref{fig:mr}(a) shows representative results from a 10 nm-thick film.
There are at least two major effects known to contribute to $\Delta \rho_{xx}$ in magnetic materials\cite{raquet2002prb}: the Lorentz magnetoresistance and the magnon-induced magnetoresistance.
Whereas the resistance quadratically increases with $H_z$ for the former, it linearly decreases with $H_z$ for the latter.
As is evident, $\Delta \rho_{xx}$ decreases almost linearly with increasing $H_z$ when the temperature is lower than $T_\mathrm{c}$, suggesting that the magnon-induced magnetoresistance is dominant\cite{raquet2002prb}. We fit the data with a linear function near zero field (from $\sim$0 to 10 kOe) to obtain the slope of $\Delta \rho_{xx}$ vs. $H_z$, which is defined as $\frac{\partial \Delta \rho_{xx}}{\partial H_z}$. 
$\frac{\partial \Delta \rho_{xx}}{\partial H_z}$ is plotted as a function of temperature for all films in Fig.~\ref{fig:mr}(b). Clearly, $\frac{\partial \Delta \rho_{xx}}{\partial H_z}$ increases as the temperature approaches $T_\mathrm{c}$, suggesting that electron-magnon scattering plays a larger role at higher temperatures.

In Fig.~\ref{fig:mr}(c-e), we plot $\frac{\partial \Delta \rho_{xx}}{\partial H_z}$ vs. $\rho_{xx}^\mathrm{m}$ to study if they are correlated. 
At the lowest temperature [Fig.~\ref{fig:mr}(c)], there is no significant correlation between the two quantities, as electron-magnon scattering is suppressed in this temperature range. As the temperature is raised[Fig.~\ref{fig:mr}(d,e)], we observe a positive correlation between the two, indicating that $\frac{\partial \Delta \rho_{xx}}{\partial H_z}$ is dependent on electron-magnon scattering.

\subsection{Anomalous Hall resistance}
\begin{figure}[H]
	\centering
	\includegraphics[width=1.0\linewidth]{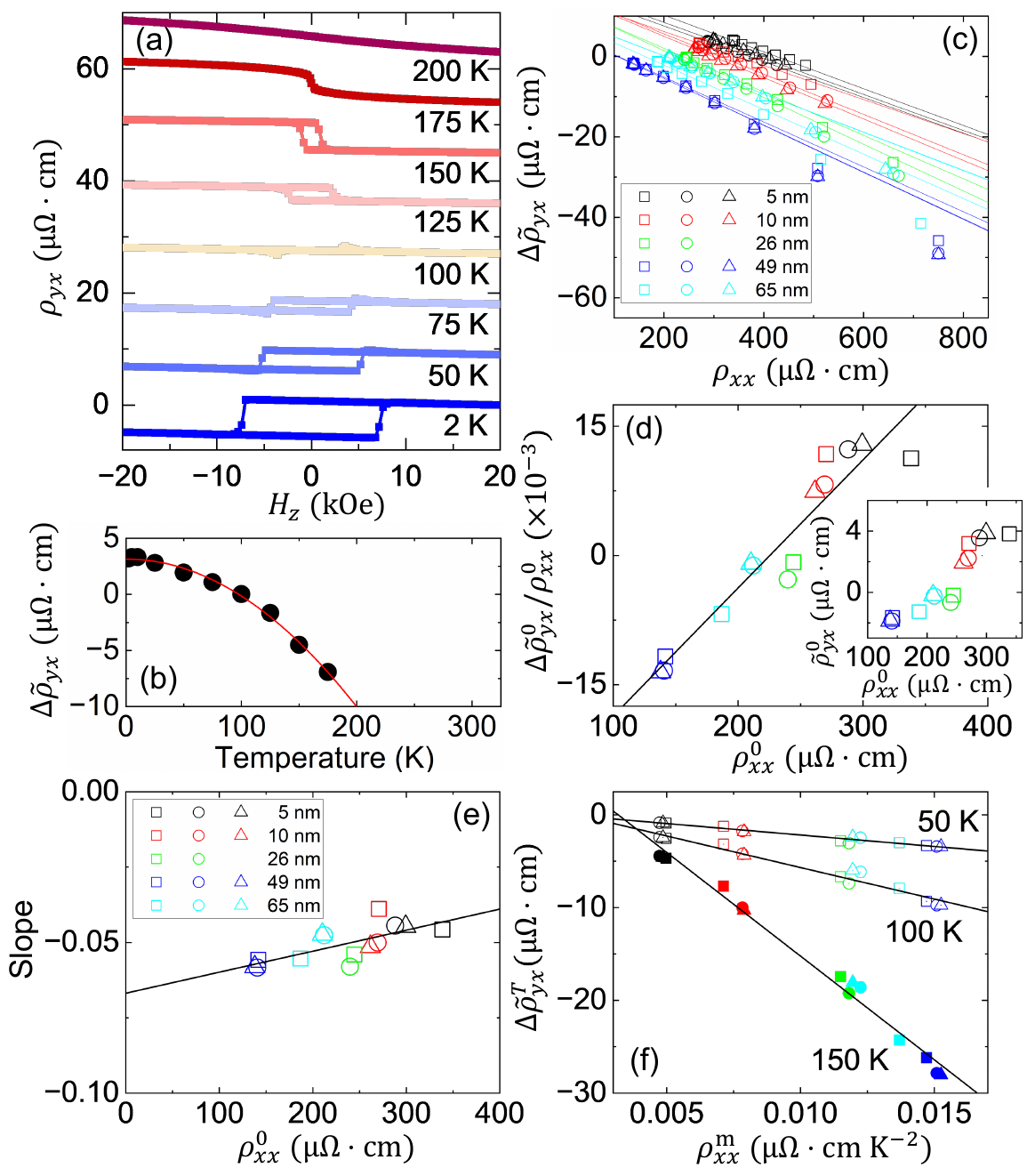}		
	\caption{\label{fig:ahe}
    \textbf{Anomalous Hall resistivity.} (a) Transverse resistivity $\rho_{yx}$, measured at different temperatures, is plotted against the out of plane magnetic field $H_z$ for a 10 nm-thick Cr$_2$Te$_3$ film. Data are shifted vertically for clarity.
    (b) Normalized anomalous Hall resistivity $\Delta \tilde{\rho}_{yx} = \Delta \rho_{yx} / \frac{M_\mathrm{s}}{M_\mathrm{s}^0}$ (black circles) of a 10 nm-thick Cr$_2$Te$_3$ film plotted against the temperature $T$. Red solid lines show a parabolic fitting to the data.
    (c) The symbols indicate the longitudinal resistivity $\rho_{xx}$ dependence of $\Delta \tilde{\rho}_{yx}$. Data displayed are from a temperature range of 2 K to 175 K. Fit to the data with a linear function is shown by the solid lines. 
    (d) $\frac{\Delta \tilde{\rho}_{yx}^0}{\rho_{xx}^0}$ plotted against the impurity scattering coefficient $\rho_{xx}^0$ (symbols). The solid line shows fit to the data using Eq.~(\ref{eq:model:ahe:zero}). The inset shows $\Delta \tilde{\rho}_{yx}^0$, i.e. $\Delta \tilde{\rho}_{yx}$ obtained at the lowest temperature, plotted as a function of $\rho_{xx}^0$. Definition of the symbols are the same as in (c): see the legend shown in (c).
	(e) The slope of the linear function used to fit the data shown in (c) plotted against $\rho_{xx}^0$ (symbols). The solid line is a linear fit to the data.
	(f) $\Delta \tilde{\rho}_{yx}^T \equiv \Delta \tilde{\rho}_{yx}-\Delta \tilde{\rho}_{yx}^0$ plotted as a function of electron-magnon scattering coefficient $\rho_{xx}^\mathrm{m}$. The open, dot center and solid symbols show data obtained at temperatures of $50, 100, 150$ K, respectively. The color of the symbols represents the film thickness whereas the symbol shape indicates results from different devices: see the legend shown in (e). The solid linear lines are guide to the eye.
	(c-f) For a given film thickness, results from a few devices are presented using different symbols.}
\end{figure}

Next, we study the transverse resistivity $\rho_{yx}$ of Cr$_2$Te$_3$.
The $H_z$ dependence of $\rho_{yx}$ of a 10 nm-thick film, measured at different temperatures, are plotted in Fig.~\ref{fig:ahe}(a). 
A clear hysteresis loop is observed at temperatures below $T_\mathrm{C}$. 
The small hump near the magnetization switching fields may be due to the topological Hall effect or the superposition of competing anomalous Hall effects with opposite signs. 
Although similar features have been observed in other systems and were attributed to the topological Hall effect, e.g. Cr$_5$Te$_6$\cite{chen2023advfuncmater}, CrTe$_2$/Bi$_2$Te$_3$\cite{zhang2021acsnano} and Cr$_2$Te$_3$/Cr$_2$Se$_3$\cite{jeon2022acsnano}, we cannot identify its origin in Cr$_2$Te$_3$ single layer film from the current data set.
The anomalous Hall resistivity $\Delta \rho_{yx}$ is obtained by subtracting the linear background found at high magnetic field and taking half the difference of the background subtracted $\rho_{yx}$ at positive and negative $H_z$.
The linear background is predominantly caused by the ordinary Hall effect: see 
Fig.~\ref{fig:transport:highfield} for the carrier density and mobility estimated from the background signal.
We normalize $\Delta \rho_{yx}$ with $\frac{M_\mathrm{s}}{M_\mathrm{s}^0}$ to exclude the temperature dependence of $M_\mathrm{s}$ from $\Delta \rho_{yx}$\cite{nagaosa2010rmp}.
The normalized anomalous Hall resistivity $\Delta \tilde{\rho}_{yx} \equiv \Delta \rho_{yx} / \frac{M_\mathrm{s}}{M_\mathrm{s}^0} $ is plotted as a function of temperature in Fig.~\ref{fig:ahe}(b). 
(See Figure~\ref{fig:transport} for the temperature dependence of $\Delta \rho_{yx}$.)
As is evident, $\Delta \tilde{\rho}_{yx}$ changes its sign at $T \sim 100$ K, a trend that has been reported in previous studies\cite{li2019acsapplnanomater, chi2023ncomm, sun2021aip, jiang2020prb}.
Similar to $\rho_{xx}$, $\Delta \tilde{\rho}_{yx}$ exhibits a $T^2$ scaling.
The red solid line in Fig.~\ref{fig:ahe}(b) shows a parabolic fitting, which agrees well with the data.
Later, we show that such $T^2$ scaling is one of the characteristics of magnon-induced skew scattering.

The scaling relation between the anomalous Hall and the longitudinal resistivities is studied to identify the origin of the anomalous Hall effect\cite{tian2009prl,hou2015prl,grigoryan2017prb}.
$\Delta \tilde{\rho}_{yx}$ is plotted against $\rho_{xx}$ in Fig.~\ref{fig:ahe}(c).
$|\Delta \tilde{\rho}_{yx}|$ increases with increasing $\rho_{xx}$, with the temperature as an implicit parameter, exhibiting a predominantly linear scaling.
In general, the scattering sources that cause the temperature dependence of the anomalous Hall resistivity can be classified into two categories, static and dynamic disorders. The former is caused by impurities and scales with $\rho_{xx}^0$, whereas the latter can be induced by magnons and phonons and is proportional to $\rho_{xx}^T \equiv \rho_{xx} - \rho_{xx}^0$.
From multi-variable scaling derived by Hou \textit{et al}.\cite{hou2015prl}, the anomalous Hall resistivity can be expressed as:
\begin{equation}
\begin{aligned}
\label{eq:model:ahe:def}
&\Delta \tilde{\rho}_{yx} = \Delta \tilde{\rho}_{yx}^\mathrm{skew} + \Delta \tilde{\rho}_{yx}^\mathrm{sj} + \Delta \tilde{\rho}_{yx}^\mathrm{int},\\
&\ \ \ \Delta \tilde{\rho}_{yx}^\mathrm{skew} = a_1 \rho_{xx}^0 + a_2 \rho_{xx}^T,\\
&\ \ \ \Delta \tilde{\rho}_{yx}^\mathrm{sj} = b_1 (\rho_{xx}^0)^2 + b_2 (\rho_{xx}^T)^2 + b_3 \rho_{xx}^0 \rho_{xx}^T,\\
&\ \ \ \Delta \tilde{\rho}_{yx}^\mathrm{int} = c \rho_{xx}^2,
\end{aligned}
\end{equation}
where $\Delta \tilde{\rho}_{yx}^\mathrm{skew}$, $\Delta \tilde{\rho}_{yx}^\mathrm{sj}$, $\Delta \tilde{\rho}_{yx}^\mathrm{int}$ are contributions from the skew scattering, the side jump and the Berry curvature effect, respectively.
$a_1$, $a_2$, $b_1$, $b_2$, $b_3$ and $c$ are the scaling coefficients.
In contrast to the note made in Ref.~\cite{hou2015prl}, here we have included a dynamic skew scattering (the $a_2$-term).

We first study the effect of static disorders on the anomalous Hall resistance.
We set $\rho_{xx}^T = 0$ and rearrange Eq.~(\ref{eq:model:ahe:def}) to obtain
\begin{equation}
\begin{aligned}
\label{eq:model:ahe:zero}
\frac{\Delta \tilde{\rho}_{yx}^0}{\rho_{xx}^0}  = a_1 + (b_1 + c) \rho_{xx}^0
\end{aligned}
\end{equation}
where $\Delta \tilde{\rho}_{yx}^0$ is $\Delta \tilde{\rho}_{yx}$ measured at the lowest temperature (2 K).
In Fig.~\ref{fig:ahe}(d), we plot $\frac{\Delta \tilde{\rho}_{yx}^0}{\rho_{xx}^0}$ as a function of ${\rho_{xx}^0}$ to determine $a_1$ and $b_1 +c$.
(As a reference, $\Delta \tilde{\rho}_{yx}^0$ vs. $\rho_{xx}^0$ is plotted in the inset.)
Data is fitted with Eq.~(\ref{eq:model:ahe:zero}): the fitted curve is shown by the solid line.
We find $a_1 \sim -0.034$ and $b_1 + c \sim 1.5 \times 10^{-4}$ $(\upmu \Omega \ \mathrm{cm})^{-1}$.
These results show that films with larger $\rho_{xx}^0$ exhibit positive $\Delta \tilde{\rho}_{yx}^0$ due to the larger contribution from the impurity induced side-jump/Berry curvature effect, i.e. the $b_1 + c$ term. This is the case for the thinner films. For the thicker films, $\Delta \tilde{\rho}_{yx}^0$ is negative since contribution from the impurity induced skew scattering ($a_1$ term) is larger.

Next, we look into the influence of dynamic disorders, which cause the temperature dependent anomalous Hall resistance.
We find that most of the data ($\Delta \tilde{\rho}_{yx}$ vs. $\rho_{xx}$) can be described by a linear function, shown by the solid lines in Fig.~\ref{fig:ahe}(c), particularly when the film thickness is small. The predominant linear dependence indicates that skew scattering (the $a_2$ term in Eq.~(\ref{eq:model:ahe:def})) contributes to the anomalous Hall effect, consistent with previous reports on Cr-Te systems\cite{cho2023nanoconv, huang2021acsnano, liu2018prb, jiang2020prb}. 
We fit the data from $2 \sim 100$ K with a linear line.
From Eq.~(\ref{eq:model:ahe:def}), the slope of the linear line is equal to $a_2 + (b_3 + 2 c) \rho_{xx}^0$.
(Note that the slope of $\Delta \tilde{\rho}_{yx}$ vs. $\rho_{xx}^T$ is the same as that of $\Delta \tilde{\rho}_{yx}$ vs. $\rho_{xx}$ since $\rho_{xx}^0$ is a constant.)
We thus plot the slope as a function of $\rho_{xx}^0$ in Fig.~\ref{fig:ahe}(e) and fit a linear function, which is shown by the solid line.
From the fitting, we find $a_2 \sim -0.067$ and $b_3 + 2 c \sim 0.7 \times 10^{-4}$ $(\mu \Omega \ \mathrm{cm} )^{-1}$.
$a_2$ is comparable in magnitude with the skew scattering coefficient in other systems\cite{vidal2011prb,gabor2015prb,meng2017pla}.

Before discussing the origin of the $a_2$ term, we comment on the other terms in the anomalous Hall resistivity.
As is evident in Fig.~\ref{fig:ahe}(c), the data deviates from the linear fitting for the thicker films at higher temperatures. 
The deviation is caused by the side jump and/or Berry curvature contributions, i.e., $b_2$ and $c$ terms in Eq.~(\ref{eq:model:ahe:def}).
(For the thinner films, because of the lower resistivity at high temperature, influence from the quadratic terms is limited.)
A previous study showed that the Berry curvature contribution in Cr$_2$Te$_3$ can vary with temperature due to thermal broadening of the Fermi surface and may depend on the film thickness via growth induced strain\cite{chi2023ncomm,song2025advfuncmater}: see 
Sec.~\ref{sec:supps:dft} and Fig.~\ref{fig:dft}
for the first principles calculations we performed as well.
The possible change of the Berry curvature contribution with temperature and thickness make it difficult to extract the coefficients $b_2$ and $c$ using Eq.~(\ref{eq:model:ahe:def}).
We therefore focus on the predominant linear term ($a_2$) and simply note that the combined contributions from the quadratic terms ($b_2$ and $c$) take a maximum of $\sim 25$\% for the thickest film near the Curie temperature, which is estimated from the deviation of $\Delta \tilde{\rho}_{yx}$ from the linear fitting.
\begin{figure}[t]
	\centering
	\includegraphics[width=1.0\linewidth]{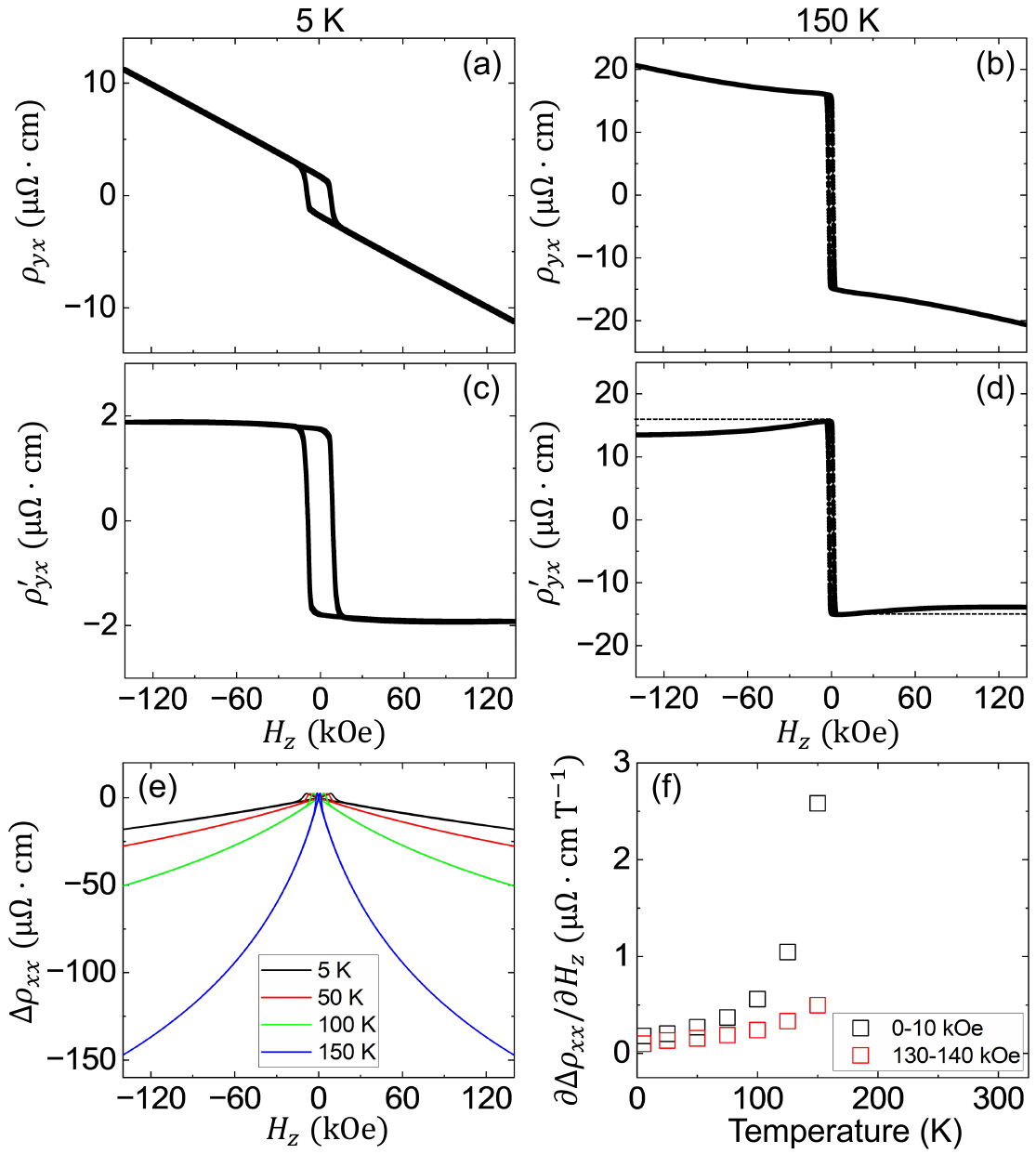}		
	\caption{\label{fig:highfield} 
	\textbf{High field longitudinal and transverse resistivities.} (a,b) Transverse resistivity $\rho_{yx}$ vs. out of plane magnetic field $H_z$ measured at 5 K (a) and 150 K (b) for a 65 nm-thick film. (c,d) $H_z$ dependence of the linear background subtracted transverse resistivity $\rho_{yx}'$. The background is determined by fitting the data in the field range of 130-140 kOe with a linear function. 
	The horizontal dotted line in (d) is a guide to the eye. 
	(e) $H_z$ dependence of the magnetoresistance $\Delta \rho_{xx}$ of a 65 nm sample measured up to 140 kOe.
(f) The slope $\partial \Delta \rho_{xx}/\partial H_z$ of the linear line fitted to $\Delta \rho_{xx}$ vs. $H_z$ in field ranges 0-10 kOe (black squares) and 130-140 kOe (red circles) are plotted as a function of temperature. 
	}
\end{figure}

To identify the source of the $a_2$ term,
we plot $\Delta \tilde{\rho}_{yx}^T \equiv \Delta \tilde{\rho}_{yx}-\Delta \tilde{\rho}_{yx}^0$ as a function of $\rho_{xx}^m$ in Fig.~\ref{fig:ahe}(f).
$\Delta \tilde{\rho}_{yx}^T$ is positively related to $\rho_{xx}^m$, particularly at higher temperatures, suggesting that magnons play a dominant role in the dynamic disorder induced scattering. 
To corroborate this observation, the $H_z$ dependence of the Hall resistance is measured up to 140 kOe for the 65 nm-thick film. 
The results obtained at measurement temperatures of 5 K and 150 K are shown in Fig.~\ref{fig:highfield}(a,b). 
At low temperature [Fig.~\ref{fig:highfield}(a)], $\rho_{yx}$ scales linearly with $H_z$ for fields outside the hysteresis loop, whereas it varies in a non-linear fashion for higher temperature [Fig.~\ref{fig:highfield}(b)]. 
($\rho_{yx}$ that scales with $H_z$ at large field is mostly caused by the ordinary Hall effect.)
To display the effect more clearly, we fit the data with a linear function in the range of $H_z \sim 130 - 140$ kOe and subtracted it from the data. 
The subtracted data, defined as $\rho_{yx}'$, are shown in Fig.~\ref{fig:highfield}(c,d).
$\rho_{yx}'$ tends to decrease as $|H_z|$ increases when the temperature is high, whereas it is nearly constant in the entire field range for lower temperature. 
Previous studies have shown that large magnetic field suppresses excitation of magnons\cite{raquet2002prb,kikkawa2015prb}. 
The reduction of $\Delta \rho_{yx}$ at large $H_z$ can therefore be attributed to decrease in magnon population.
These results thus support the notion that electron-magnon scattering contributes to the anomalous Hall resistance at higher temperatures. 
We note that the saturation of $\rho_{yx}'$ in Fig.~\ref{fig:highfield}(d) is caused by the linear background subtraction process. Measurements at even larger magnetic field are needed to determine the saturation field.
Indeed, previous studies showed that suppression of magnon induced effects requires magnetic field of the order of a few hundreds of kOe\cite{raquet2002prb}.

Suppression of electron-magnon scattering by magnetic field can also be found in the magnetoresistance measured at higher magnetic field.
Figure~\ref{fig:highfield}(e) shows the magnetoresistance $\Delta \rho_{xx}$ measured up to 140 kOe for the 65 nm-thick film.
The slope of $\Delta \rho_{xx}$ vs. $H_z$ clearly changes with $H_z$ at higher temperatures.
Note that the slope represents the strength of electron-magnon scattering: see the discussion pertaining to Fig.~\ref{fig:mr}.
We plot the temperature dependence of the slope $\partial \Delta \rho_{xx}/\partial H_z$ at lower magnetic field (0-10 kOe) and higher magnetic field (130 -140 kOe) in Fig.~\ref{fig:highfield}(f).
The former is significantly larger than the latter when the temperature is high, indicating that magnon excitation is suppressed at larger magnetic field.
These results strongly suggest that the source of the $a_2$ term is magnon-induced skew scattering.

Based on these results, we discuss the reason behind the sign change of the anomalous Hall resistance with temperature in Cr$_2$Te$_3$. At the lowest temperature, $\Delta \tilde{\rho}_{yx}^0$ is governed by static disorder (impurity) scattering $\rho_{xx}^0$. 
The linear ($a_1$) and quadratic ($b_1+c$) terms have opposite sign. 
For the thinner films (with larger $\rho_{xx}^0$), the net contribution is positive since the quadratic term dominates. 
In contrast, $\Delta \tilde{\rho}_{yx}^0$ is negative for the thicker films (with smaller $\rho_{xx}^0$) as the linear term is dominant. With increasing temperature, contribution from the magnon induced skew scattering ($a_2$), which is negative, increases and takes over, resulting in a sign change of $\Delta \tilde{\rho}_{yx}$ only for the thinner films. 

\section{\label{sec:model}Model calculations}
Finally, we discuss the microscopic origin of the magnon-induced skew scattering in Cr$_2$Te$_3$. 
The anomalous Hall resistivity $\Delta \tilde{\rho}_{yx}$ that originates from magnon-induced skew scattering is obtained by including the effect of the spin-orbit coupling in a $p$-$d$ exchange interaction. 
The Hamiltonian that describes the electron-magnon skew scattering has been proposed as\cite{irkhin2021condmat}:
\begin{equation}
\begin{aligned}
\label{eq:model:H}
\mathcal{H}_{pd} 
&= i \lambda J_{pd} a_0^2 \sum_{\bm{k}, \bm{k}'} \left( \bm{k} \times \bm{k}' \right) \cdot \left( \delta \bm{S} \right)_{\bm{k} - \bm{k}'} c_{\bm{k}}^\dagger c_{\bm{k}'},
\end{aligned}
\end{equation}
where $\lambda$ is a dimensionless parameter that characterizes the spin-orbit coupling, $J_{pd}$ represents the $p$-$d$ exchange interaction between the Te 5$p$ conduction electrons and the Cr 3$d$ localized moments, $a_0^3$ is the volume per localized spin, $\bm{k}$ and $c_{\bm{k}}$ are the electron wave vector and annihilation operator, $\bm{S}$ represents the localized spin ($S = |\bm{S}|=3/2$ for Cr) and $\delta \bm{S}$ is its deviation from the equilibrium direction due to magnon excitation.
Assuming a free electron like band with exchange splitting, the anomalous Hall resistivity $\Delta \tilde{\rho}_{yx}^\mathrm{cal}$ is calculated by considering the process shown in Fig.~\ref{fig:diagram}
, as
\begin{equation}
\begin{aligned}
\label{eq:model:rhoAH}
\Delta \tilde{\rho}_{yx}^\mathrm{cal} 
&= - \Xi_{yx} \frac{\lambda a_0^2 J_{pd}^3 S}{8 \hbar^2 e^2 v_\mathrm{F}^3} \left( \frac{k_\mathrm{B} T}{A_\mathrm{ex}} \right)^2,
\end{aligned}
\end{equation}
where $v_\mathrm{F}$ is the Fermi velocity and $A_\mathrm{ex}$ is the exchange stiffness parameter.
($\hbar$ is the reduced Planck constant, $e$ is the elementary charge and $k_\mathrm{B}$ is the Boltzmann constant.)
Note that this process was not considered in Ref.~\cite{hou2015prl}.
The longitudinal resistivity $\rho_{xx}^\mathrm{cal}$ due to electron-magnon scattering is given by\cite{irkhin2021condmat}
\begin{equation}
\begin{aligned}
\label{eq:model:rhoxx}
\rho_{xx}^\mathrm{cal} 
&= \Xi_{xx} \frac{m^2 J_{pd}^2 S}{(2 \pi)^3 \hbar^3 e^2 n^2 a_0^3} \left( \frac{k_\mathrm{B} T}{A_\mathrm{ex}} \right)^2,
\end{aligned}
\end{equation}
where $\Xi_{yx}$ [in Eq.~(\ref{eq:model:rhoAH})] and $\Xi_{xx}$ [in Eq.~(\ref{eq:model:rhoxx})] are constants of order unity and weakly dependent on temperature (see 
Fig.~\ref{fig:calc}). 
$m$ is the effective electron mass and $n$ is the electron density.
See Sec.~\ref{sec:supps:model} 
for the outline of the derivation of Eqs.~(\ref{eq:model:rhoAH}) and (\ref{eq:model:rhoxx}), details of $\Xi_{yx}$ and $\Xi_{xx}$ and the Feynman diagram 
Fig.~\ref{fig:diagram}) 
used to calculate the anomalous Hall conductivity due to magnon-induced skew scattering.

Both $\Delta \tilde{\rho}_{yx}^\mathrm{cal}$ and $\rho_{xx}^\mathrm{cal}$ scale with the temperature $T$ quadratically, in agreement with the experiments: see Figs.~\ref{fig:rhoxx}(b) and \ref{fig:ahe}(b).
Substituting the parameters listed in 
Table~\ref{table:supps:para}
 suitable for Cr$_2$Te$_3$, we obtain $\Delta \tilde{\rho}_{yx}^\mathrm{cal} \sim - 22$ $\mu \Omega \cdot$cm and $\rho_{xx}^\mathrm{cal} \sim 329$ $\mu \Omega \cdot$cm at $T=200$ K from Eqs.~(\ref{eq:model:rhoAH}) and (\ref{eq:model:rhoxx}). Here we adjusted $\lambda$, the dimensionless parameter that characterizes the spin-orbit interaction, to match the value of $a_2 = -0.067$ obtained in the experiments with $\Delta \tilde{\rho}_{yx}^\mathrm{cal} / (\rho_{xx}^\mathrm{cal})$.
Note that $J_{pd} = -0.1$ eV is estimated from the band structure of Cr$_2$Te$_3$ using first principles calculations. $J_{pd}$ is normally negative when the exchange coupling is between conduction electrons and localized moments that belong to different elements.
From Eq.~(\ref{eq:model:rhoAH}), we find $\lambda$ must be negative in order to account for the negative $\Delta \tilde{\rho}_{yx}$ found in the experiments (see e.g. Fig.~\ref{fig:ahe}(b)).
The strength of the $p-d$ exchange interaction, characterized by $\lambda J_{pd}$, is $\sim 0.084$ eV, which is similar in magnitude with the atomic spin-orbit coupling of Te\cite{doi1970jpsj,furukawa2017ncomm,sakano2020prl}.
We thus infer that the predominant magnon-induced skew scattering in Cr$_2$Te$_3$ is primarily induced by the large spin-orbit coupling of Te.
A microscopic model that takes into account the band structure of host material (here Cr$_2$Te$_3$) is required to clarify the relation between the atomic spin orbit coupling and $\lambda$.

\section{\label{sec:conclusion}Conclusion}
We have studied the longitudinal and anomalous Hall resistivities of Cr$_2$Te$_3$ thin films. We find a quadratic temperature dependence of the longitudinal resistivity below the Curie temperature, suggesting that electron-magnon scattering is one of the major sources of the resistivity. This is corroborated by a linear magnetoresistance found against the out-of-plane magnetic field.
The anomalous Hall resistivity includes two major contributions: temperature dependent and temperature independent terms.
The former exhibits a predominant linear dependence with the longitudinal resistivity and is positively correlated with the electron-magnon scattering coefficient of the longitudinal resistivity, suggesting that magnon induced skew scattering significantly contributes to the anomalous Hall effect in Cr$_2$Te$_3$.
We also find the anomalous Hall resistivity at higher temperature is suppressed by large magnetic field, supporting the scenario that the skew scattering originates from collision with magnons.
For the latter (i.e. the temperature independent term), we find that the contribution from the impurity induced side-jump and/or the Berry curvature effect possess the opposite sign with that of magnon-induced skew scattering.
%
The sign change of the anomalous Hall resistivity with temperature is thus accounted for by the competition between the two effects.
Using model calculations, we show that the $p$-$d$ type exchange interaction modified by spin-orbit coupling can account for the significant electron-magnon scattering contribution to the anomalous Hall effect in Cr$_2$Te$_3$.
These results suggest that magnon-induced skew scattering plays a significant role in the anomalous Hall effect in layered ferromagnets with heavy elements, a class of 2D materials that are of current interest.

\begin{acknowledgments}
This work was partly supported by JSPS KAKENHI (Grant Numbers 21H01608, 22K14290, 22K18935, 23H00176, 23H05463), JST CREST (JPMJCR19T3), MEXT Initiative to Establish Next-generation Novel Integrated Circuits Centers (X-NICS) and Cooperative Research Project Program of RIEC, Tohoku University. Computations were performed on a Numerical Materials Simulator at NIMS. Y.W. is supported by the GSGC program of University of Tokyo, S.W. thanks JST SPRING GX, Grant Number JPMJSP2108.
\end{acknowledgments}

\section{Methods}
\subsection{\label{sec:method:sample}Sample preparation}
Cr$_2$Te$_3$ films were grown on sapphire (0001) or MgO (001) substrates in a commercial molecular beam epitaxy (MBE) system. 
The substrates were pre-annealed in vacuum at 650 $^{\mathrm{o}}$C for 150 min for degassing. After degassing, the substrate temperature $T_\mathrm{s}$ was set to $\sim 380$ $^{\mathrm{o}}$C. Tellurium and chromium were co-evaporated using a Knudsen cell for Te and an electron beam gun for Cr. To form Cr$_2$Te$_3$, the flux ratio of Te over Cr was kept to $\sim$20 or higher. The growth rate of Cr$_2$Te$_3$ was monitored using a quartz crystal spectrometer and its typical value is $\sim$0.01 nm/s. Cr$_2$Te$_3$ thin films with different thicknesses (5, 10, 26, 49 and 65 nm) were grown. The thickness of the films were determined using X-ray reflectivity measurements. A 5 nm-thick Te or 3 nm-thick Ti capping layer was deposited at room temperature to protect the Cr$_2$Te$_3$ film from oxidation. We find the type of substrate (sapphire vs. MgO) and the capping layer material (Te vs. Ti) have little influence on the magnetic and transport properties of the films. 

\subsection{\label{sec:method:meas}Sample characterization}
Magnetic properties of the films were examined by superconducting quantum interference diffractometer (SQUID).
The specimen for cross-sectional transmission electron microscopy (TEM) studies was prepared by using a focused ion beam (FIB). High-angle annular dark-field scanning transmission electron microscopy (HAADF-STEM) observation, nanobeam electron diffraction, and energy dispersive x-ray spectroscopy (EDS) analysis were carried out using a commercial system.

To study the transport properties, the films were patterned into Hall bars using optical lithography and Ar ion milling. Contact electrodes, made of $\sim$50 nm thick conducting materials, were patterned on the Hall bars using standard liftoff technique. The electrodes were deposited via RF magnetron sputtering. 
The width $w$ and length $L$ of the Hall bar channel are 10 $\upmu$m and 25 $\upmu$m, respectively. 
Measurements were performed using a physical property measurement system (PPMS). A current $I_x$ of 100 $\upmu$A was applied to the sample and the longitudinal voltage $V_{xx}$ and the transverse voltage $V_{yx}$ were measured. The longitudinal and transverse resistivities are obtained from the relations $\rho_{xx} = \frac{V_{xx}}{I_x} \frac{w t}{L}$ and $\rho_{yx} = \frac{V_{yx} t}{I_x}$, respectively, where $t$ is the thickness of the film. See 
Sec.~\ref{sec:supps:trans} and Fig.~\ref{fig:measurement}
for the details of the configuration used to measure the longitudinal and transverse resistivities.

\clearpage
\section{Supplementary Material}
\subsection{\label{sec:supps:structure} \large{S\lowercase{tructural characterization}}}
\begin{figure}[h]
	\includegraphics[width=1.0\linewidth]{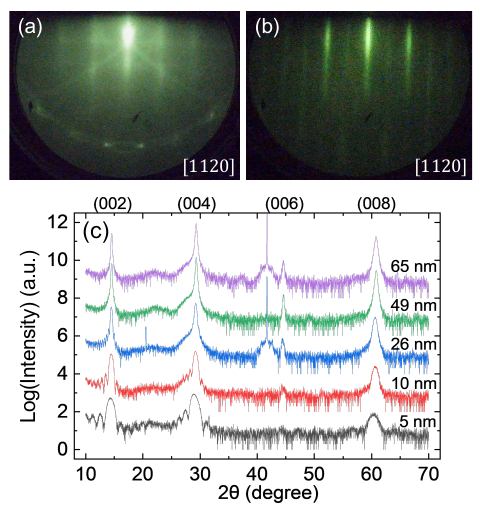}
	\caption{\label{fig:rheed} \textbf{Structural properties.} (a,b) RHEED pattern of the post-annealed sapphire substrate (a) and upon deposition of a 65 nm-thick Cr$_2$Te$_3$ film. (c) XRD spectra ($\theta - 2\theta$ scan) of the Cr$_2$Te$_3$ films with different thicknesses.}
\end{figure}

\normalsize
Reflective high energy electron diffraction (RHEED) was used to in-situ monitor the film growth. Figure~\ref{fig:rheed}(a,b) shows the RHEED pattern of the post-annealed substrate and a 65 nm-thick Cr$_2$Te$_3$ film. Kikuchi lines and the second diffraction spots are found in the RHEED pattern of the substrate, suggesting that the substrate surface is well defined before deposition of Cr$_2$Te$_3$. The RHEED pattern of Cr$_2$Te$_3$ shows streaky vertical lines, a signature of flat surface.
The X-ray diffraction (XRD) spectra of the films with different thicknesses are presented in Fig.~\ref{fig:rheed}(c). The peak position and intensity found in the spectra are consistent with (001) oriented growth of Cr$_2$Te$_3$. The lattice constant along the film normal changes with the film thickness: see Fig.~\ref{fig:dft}(c) for the details.


\subsection{\label{sec:supps:trans} \large{M\lowercase{agnetic and transport properties}}}
\normalsize
\begin{figure}[h]
	\centering
	\includegraphics[width=1.0\linewidth]{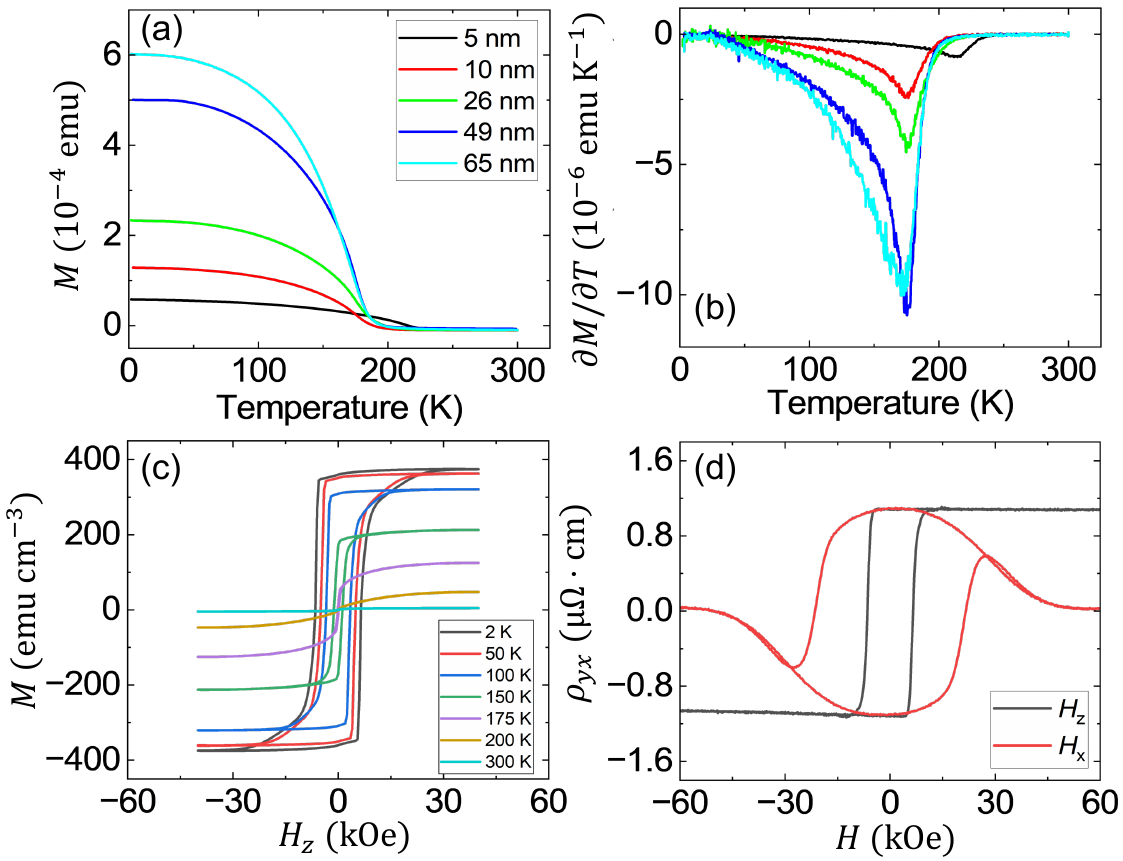}		
	\caption{\label{fig:vsm} \textbf{Magnetic properties.} (a,b) Temperature dependence of magnetic moment $M$ (a) and its temperature derivative $\partial M/\partial T$ (b). $M$ is measured under a constant out-of-plane magnetic field of $\sim$500 Oe. The Curie temperature   $T_\mathrm{C}$ is estimated from the temperature at which $\partial M/\partial T$ takes a minimum. (c) Out-of-plane magnetic field $H_z$ dependence of the magnetic moment per unit volume, measured at different temperatures, for the 65 nm-thick film. (d) Anomalous Hall resistivity $\rho_{yx}$ plotted as function of out-of-plane field $H_z$ (black line) and in-plane field $H_x$ (red line) for a 4 nm-thick film measured at 50 K. A linear background due to the ordinary Hall effect is subtracted from the data.}
\end{figure}
The magnetic properties of the films were studied using a commercial vibrating sample magnetometry (VSM). To estimate the Curie temperature $T_\mathrm{C}$, the temperature dependence of the magnetic moment was measured. Figure~\ref{fig:vsm}(a) shows the magnetic moment $M$ measured under a fixed magnetic field ($\sim 500$ Oe) applied along the film normal while the temperature was swept from room temperature to 2 K (field cooling). Here the temperature is scanned in steps of $\sim$0.1 K while the magnetization is measured (note that the data presented in Fig.~\ref{fig:rhoxx}(a) of the main text was obtained by sweeping the magnetic field at fixed temperatures).  The temperature derivative of $M$ ($\partial M/\partial T$) is plotted against the temperature in Fig.~\ref{fig:vsm}(b). As is evident, $\partial M/\partial T$ takes a minimum, which corresponds to $T_\mathrm{C}$. We find that $T_\mathrm{C}$ is close to 175 K except for the 5 nm-thick film which has a $T_\mathrm{C}$ of $\sim$215 K.
\begin{figure}[h]
	\centering
	\includegraphics[width=0.9\linewidth]{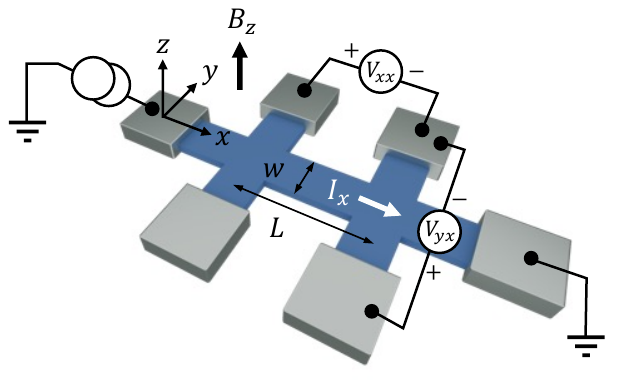}		
	\caption{\label{fig:measurement} \textbf{Schematic of the device and the measurement setup.} Illustration of the setup to measure the longitudinal and transverse resistivities.}
\end{figure}
The out-of-plane magnetic field $H_z$ dependence of $M$ for the 65 nm-thick film, obtained at different temperatures, is presented in Fig.~\ref{fig:vsm}(c). Square hysteresis loops with non-zero remanence are found for temperatures well below $T_\mathrm{C}$, suggesting that the magnetic easy axis points along the film normal. Similar loops are obtained for the other films studied at low temperatures. 

Figure~\ref{fig:measurement} shows the coordinate axis and the setup used to measure the longitudinal and transverse resistivities. Current $I_x$ is passed along $+x$, magnetic field $H_z$ is applied along $+z$, and the longitudinal and transverse voltages $V_{xx}$ and $V_{yx}$ are measured using the probe configuration described in Fig.~\ref{fig:measurement}. The longitudinal and transverse resistivities are obtained from the relations $\rho_{xx} = \frac{V_{xx}}{I_x} \frac{w t}{L}$ and $\rho_{yx} = \frac{V_{yx} t}{I_x}$, respectively, where $t$ is the thickness of the film.  
\begin{figure}[b]
	\centering
	\includegraphics[width=1.0\linewidth]{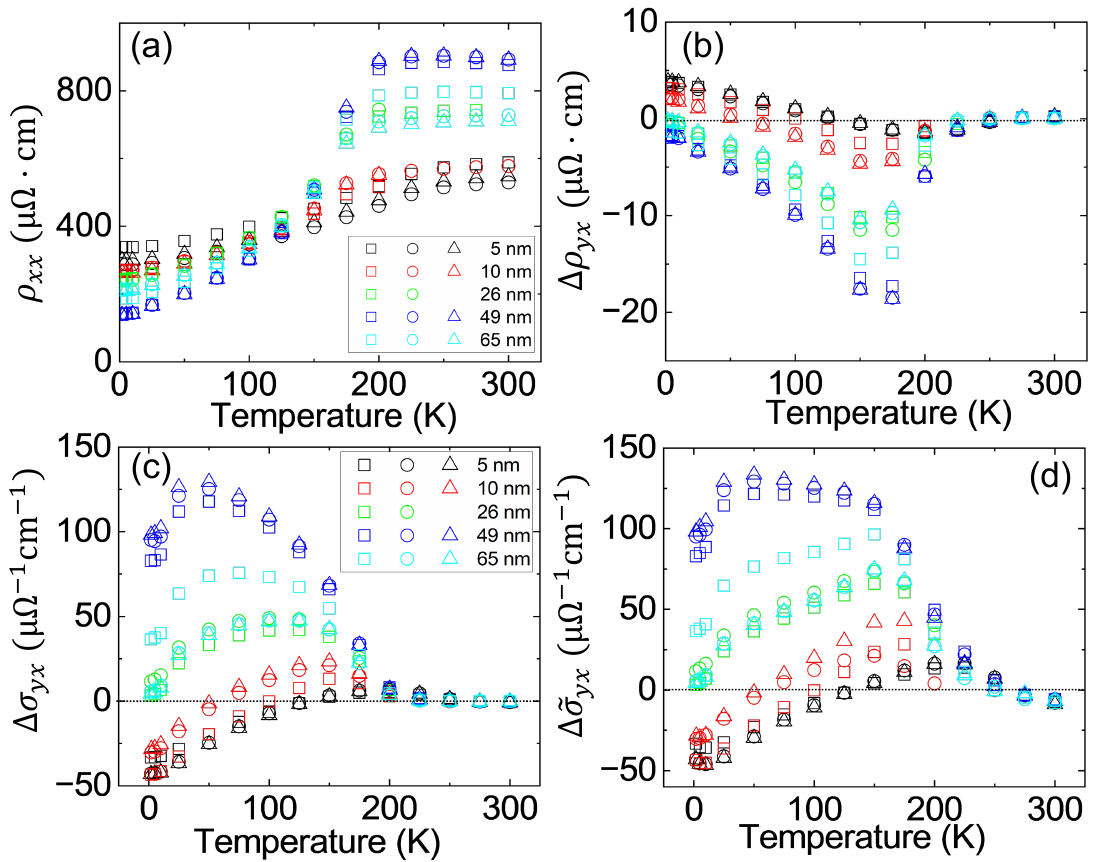}		
	\caption{\label{fig:transport} \textbf{Longitudinal and transverse resistivities.} (a-d) The temperature dependence of the longitudinal resistivity $\rho_{xx}$ (a), the anomalous Hall resistivity $\Delta \rho_{yx}$ (b), the anomalous Hall conductivity $\Delta \sigma_{yx} = - \frac{\Delta \rho_{yx}}{\rho_{xx}^2}$ (c) and the normalized anomalous Hall conductivity $\Delta \tilde{\sigma}_{yx} = \Delta \sigma_{yx} / \frac{M_\mathrm{s}}{M_\mathrm{s}^0}$ (d) for Cr$_2$Te$_3$ films with different thicknesses. Results from a few devices are presented for a given film thickness.}
\end{figure}

We first estimate the magnetic anisotropy field $H_\mathrm{K}$. Here we show results from a 4 nm-thick film.
The deposition condition, structural and transport properties of this film are similar to those of the 5-nm thick film presented in the main text. 
The in-plane $H_x$ and out-of-plane $H_z$ magnetic field dependence of the anomalous Hall resistivity is presented in Fig.~\ref{fig:vsm}(d). The measurement temperature is 50 K.
The out-of-plane loop shows a square hysteresis whereas the in-plane loop closes at a magnetic field of $H_\mathrm{K}\sim$50 kOe. 
From the data presented in Fig.~\ref{fig:rhoxx}(a) of the main text, the saturation magnetization reads $M_\mathrm{s} \sim 400$ emu cm$^{-3}$. The magnetic anisotropy energy density $K_\mathrm{eff} = \frac{1}{2} M_\mathrm{s} H_\mathrm{K}$ is thus $\sim 1 \times 10^7$ erg cm$^{-3}$. 

The temperature dependence of the longitudinal resistivity $\rho_{xx}$ and the anomalous Hall resistivity $\Delta \rho_{yx}$ are plotted in Fig.~\ref{fig:transport}(a,b) for films with different thicknesses. 
The anomalous Hall conductivity $\Delta \sigma_{yx} = - \frac{\Delta \rho_{yx}}{\rho_{xx}^2}$ and its normalized value $\Delta \tilde{\sigma}_{yx} = \Delta \sigma_{yx} / \frac{M_\mathrm{s}}{M_\mathrm{s}^0}$ are plotted as a function of temperature in Figs.~\ref{fig:transport}(c) and \ref{fig:transport}(d), respectively. 
The symbol colors represent data from samples with different thicknesses. Two to three samples are measured for a given film thickness and the results are shown using different symbols.

\begin{figure}[h]
	\centering
	\includegraphics[width=1.0\linewidth]{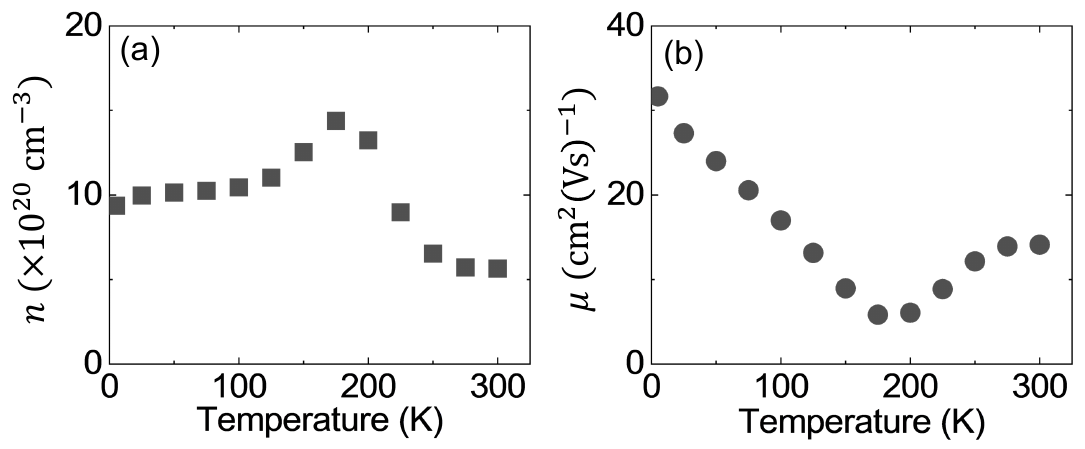}		
	\caption{\label{fig:transport:highfield} \textbf{Carrier density and mobility.} (a,b) The temperature dependence of carrier density $n$ (a) and mobility $\mu$ (b) for 65 nm-thick Cr$_2$Te$_3$ film. Positive $n$ in (a) indicates that the majority carrier type is holes.}
\end{figure}

The carrier density $n$ is estimated using the change in the transverse resistivity $\rho_{yx}$ with $H_z$. Here we use a setup that can apply maximum 140 kOe to minimize contributions from the anomalous hall effect on the ordinary Hall effect. 
We calculate the slope of $\rho_{yx}$ vs. $H_z$ to estimate $n$. Using the relation $\mu = \frac{1}{\rho_{xx} e n}$, the mobility $\mu$ is obtained. $n$ and $\mu$ are plotted as a function of measurement temperature in Figs.~\ref{fig:transport:highfield}(a) and \ref{fig:transport:highfield}(b), respectively.
Since contributions from the anomalous Hall effect cannot be completely excluded even with application of a field of 140 kOe, 
the results presented in Fig.~\ref{fig:transport:highfield}(a,b) provide an upper limit on $n$ and a lower limit on $\mu$.

\subsection{\label{sec:supps:dft} \large{F\lowercase{irst principles calculations}}}
\begin{figure}[b]
	\centering
	\includegraphics[width=1.0\linewidth]{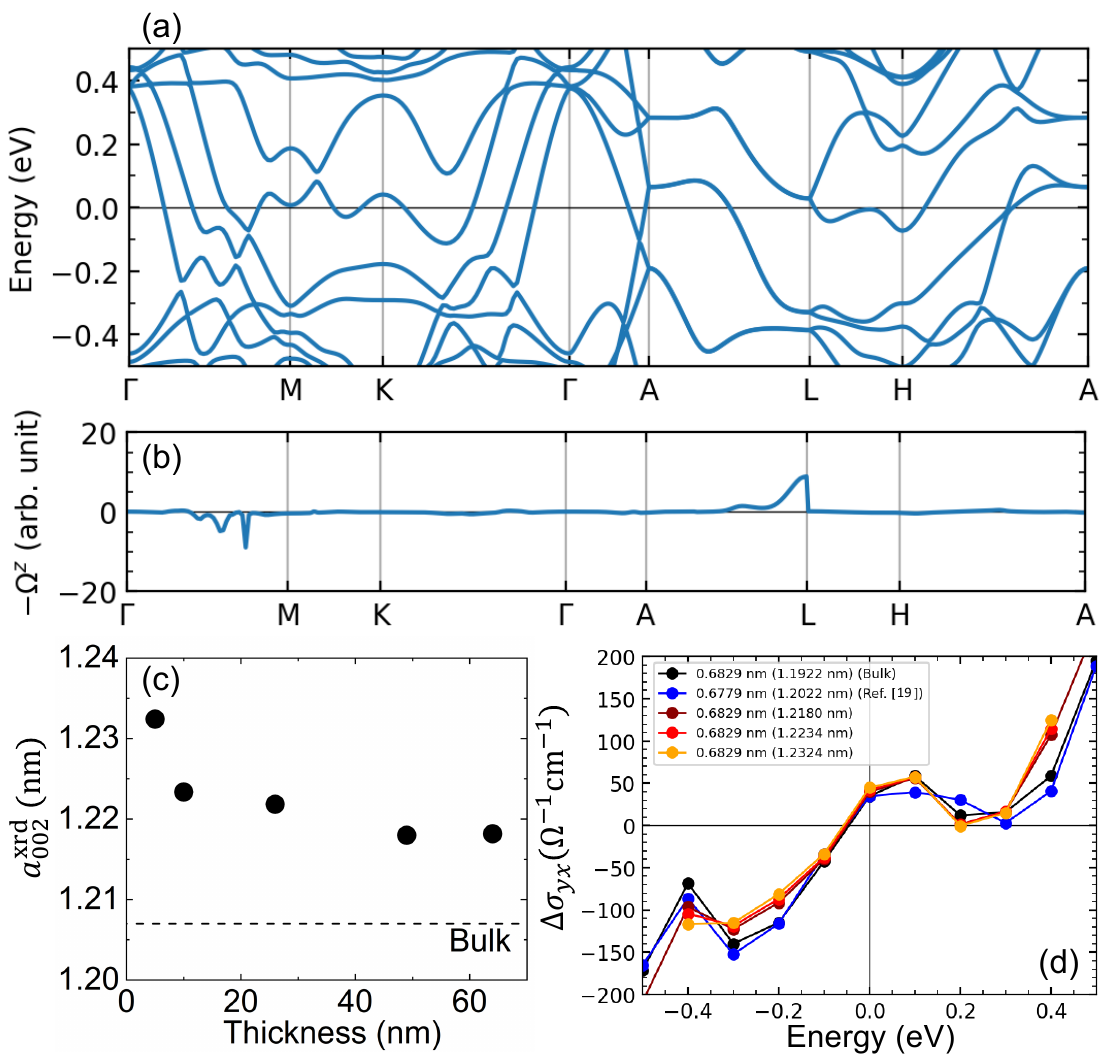}		
	\caption{\label{fig:dft} \textbf{Dispersion relation and calculated anomalous Hall conductivity.} (a,b) Calculated band structure of Cr$_2$Te$_3$ (a) and the corresponding Berry curvature $\Omega^z$ at the Fermi energy (b) plotted along high symmetry lines of the Brillouin zone. (c) Cr$_2$Te$_3$ lattice constant $a_{002}^\mathrm{xrd}$ along the $c$-axis obtained from the XRD spectra presented in Fig.~\ref{fig:rheed}(c). Calculated anomalous Hall conductivity $\Delta \sigma_{yx}$ obtained by integrating the Berry curvature across the entire Brillouin zone. Symbols indicate calculation results using different lattice constants. We denote the in-plane lattice constant with $a_{100}$. Black circles: $a_{100} = 0.6829$ nm, $a_{002} = 1.1922$ nm (same for (a)), blue circles: $a_{100} = 0.6779$ nm, $a_{002} = 1.2022$ nm, dark-red circles: $a_{100} = 0.6829$ nm, $a_{002} = 1.2180$ nm, red circles: $a_{100} = 0.6829$ nm, $a_{002} = 1.2234$ nm, orange circles: $a_{100} = 0.6829$ nm, $a_{002} = 1.2324$ nm. 
	}
\end{figure}

\normalsize
Contributions from the Berry curvature is calculated using first principles calculations. From the XRD spectra, we find the lattice is strained as the film is grown on a single crystalline substrate and the strain depends on the film thickness. 
The lattice constant $a_{002}$ along the $c$-axis (i.e. distance between the (002) planes), obtained from the XRD spectra and referred to as $a_{002}^\mathrm{xrd}$ hereafter, is plotted as a function of film thickness in Fig.~\ref{fig:dft}(c). As is evident, $a_{002}^\mathrm{xrd}$ decreases with film thickness. The dashed line shows $a_{002}$ of bulk Cr$_2$Te$_3$. $a_{002}^\mathrm{xrd}$ tends to approach $a_{002}$ of the bulk phase as the thickness is increased. 

The calculated band structure of Cr$_2$Te$_3$ and the corresponding Berry curvature $\Omega^z$ from the occupied band at each $k$ point
are plotted along the high symmetry lines in Figs.~\ref{fig:dft}(a) and \ref{fig:dft}(b). 
A large positive Berry curvature peak is found near the L point, which is responsible for the anomalous Hall effect in bulk Cr$_2$Te$_3$.
The anomalous Hall conductivity $\Delta \sigma_{yx}$ is calculated by integrating the Berry curvature over the entire Brillouin zone.
In Fig.~\ref{fig:dft}(d), we show $\Delta \sigma_{yx}$ as a function of the Fermi energy.
Symbols indicate results when the lattice constant is varied.
The blue circles represent $\Delta \sigma_{yx}$ calculated for Cr$_2$Te$_3$ with bulk lattice parameters, the purple circles show the results when the lattice constant reported in Ref.~\cite{chi2023ncomm} is used. 
The orange, green and red circles display $\Delta \sigma_{yx}$ when the in-plane lattice constant is fixed to that of the bulk but the out of plane lattice constant is set to $a_{002}^\mathrm{xrd}$.
In all cases, we find $\Delta \sigma_{yx}$ is positive near the Fermi level and its magnitude is $\sim$35 to $\sim$45 $(\Omega \ \mathrm{cm})^{-1}$ [$\Delta \sigma_{yx}$ reported in Ref.~\cite{chi2023ncomm} was $\sim-12.7$ ($\Omega \ \mathrm{cm})^{-1}$].
(With regard to the sign of anomalous Hall conductivity, we calculated $\Delta \sigma_{yx}$ for bcc-Fe as a reference: the sign of $\Delta \sigma_{yx}$ is also positive for bcc-Fe and its magnitude is $480$ $(\Omega \ \mathrm{cm})^{-1}$.)
Note that positive $\Delta \sigma_{yx}$ (or negative $\Delta \rho_{yx}$) is consistent with the experiments, i.e, the quadratic components of the dynamic disorder induced scattering terms ($b_1 + c$): see Fig.~\ref{fig:ahe}.
For Cr$_2$Te$_3$, $\Delta \sigma_{yx}$ changes its sign if the Fermi level becomes 0.06 eV lower than its equilibrium phase.
This applies to all the cases studied. 
The effect of lattice strain, in contrast, is small particularly when the Fermi level is close to that of the equilibrium phase (near zero energy level).

\subsection{\label{sec:supps:model} \large{M\lowercase{odel calculations}}}
\normalsize
Following Ref.~\cite{irkhin2021condmat}, we consider the electron-magnon interaction derived from the $s$-$d$ type exchange interaction mediated by spin-orbit coupling (SOC), and calculate the longitudinal resistivity $\rho_{xx}$ and the anomalous Hall resistivity $\rho_{yx}$. 
Here, the model is called the $p$-$d$ model since the first-principles calculation predicts localized Cr-$d$ moments 
and itinerant Te-$p$ electrons.

The Hamiltonian is given by 
\begin{equation}
\begin{aligned}
 H  =& \int d{\bm r}  \left\{ \frac{\hbar^2}{2m} |\nabla c |^2   + V_{\rm imp} c^\dagger c 
   + J (\nabla {\bm S})^2   \right\}\\
   &- J_{pd}   \int d{\bm r}  \, {\bm S} \cdot (c^\dagger {\bm \sigma} c\,)  
      +  H_{\rm so} ,    
\label{eq:Htot}
\end{aligned}
\end{equation}
where $c$ ($c^\dagger$) is the electron annihilation (creation) operator, $V_{\rm imp}$ is the impurity potential, 
$J$ is the exchange interaction among the Cr-$d$ moments, $J_{pd}$ is the $p$-$d$ exchange interaction constant between the Te 5$p$ conduction electrons and the Cr 3$d$ localized moments, $\bm{S}$ is the localized spin ($|\bm{S}|=3/2$ for Cr), $m$ is the effective electron mass, and $H_{\rm so}$ represents the effects of SOC on the $p$-$d$ exchange interaction. 
Writing ${\bm S}({\bm r},t) = S \hat z + \delta {\bm S}({\bm r},t)$, 
with small deviations $\delta {\bm S}({\bm r},t)$, we have 
$H = H_{\rm e} + H_{\rm m} + H_{\rm em} + H_{\rm so}$,  
\begin{align}
 H_{\rm e} &=  \sum_{\bm k} \varepsilon_{{\bm k} \sigma} c_{{\bm k} \sigma}^\dagger c_ {{\bm k} \sigma}^{\phantom{\dagger}}   + H_{\rm imp} , 
\\ 
 H_{\rm m} &=  \sum_{\bm q} \omega_{\bm q} a_{\bm q}^\dagger a_ {\bm q}^{\phantom{\dagger}} , 
 \\
 H_{\rm em} &= - J_{pd} \sum_{{\bm k}, {\bm k}'} (\delta {\bm S})_{{\bm k} - {\bm k}'} \cdot 
      ( c_{\bm k}^\dagger {\bm \sigma} c_ {{\bm k}'}^{\phantom{\dagger}} )  , 
\\
H_{\rm so} &=  i \lambda J_{pd} a_0^2 
    \sum_{{\bm k}, {\bm k}'} ({\bm k} \times {\bm k}') \cdot (\delta {\bm S})_{{\bm k} - {\bm k}'} \, 
       c_{\bm k}^\dagger c_ {{\bm k}'}^{\phantom{\dagger}} ,     
\label{eq:Hsd}
\end{align}
with $\varepsilon_{{\bm k} \sigma} = \hbar^2 k^2/2m - \sigma J_{pd} S$ 
($\sigma = \uparrow, \downarrow$ or $\pm 1$), $\omega_{\bm q} = J S^2 q^2$ and $a_{\bm{q}}$ ($a_{\bm{q}}^\dagger$) is the magnon annihilation (creation) operator and $a_0^3$ is the volume per localized spin.
$H_{\rm em}$ is the ordinary electron-magnon coupling and $H_{\rm so}$ may arise in the presence of SOC  
($\lambda$ is a dimensionless parameter)\cite{holstein1940pr}.
$H_{\rm so}$ was derived in Ref.~\cite{irkhin2021condmat} from the direct exchange interaction 
with the SOC of $d$ spins taken into account. 
For Cr$_2$Te$_3$, we suppose the SOC of Te $p$ electrons plays important roles to produce $H_{\rm so}$. 
In $H_{\rm em}$ and $H_{\rm so}$, we express $\delta {\bm S}$ using the magnon operators\cite{holstein1940pr}.
\begin{figure}[tb]
 \includegraphics[width=0.6\linewidth]{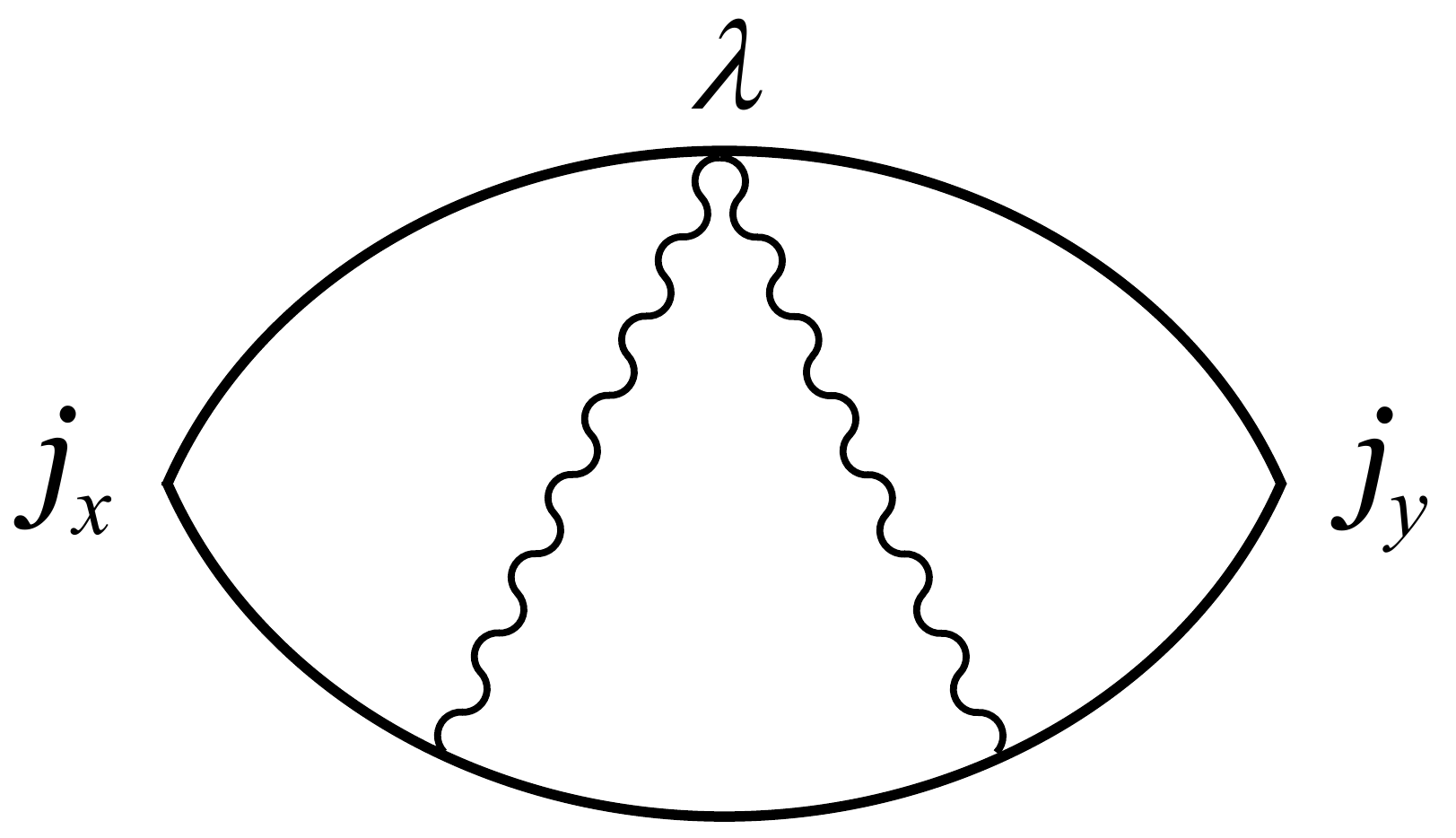}
\caption{\label{fig:diagram} \textbf{Feynman diagram.} Feynman diagram for the anomalous Hall conductivity due to magnon-induced skew scattering. The solid (wavy) lines represent electron (magnon) propagators, and $j_x$ and $j_y$ indicate current vertices.}
\end{figure}

\begin{figure*}[tb]
 \centering
 \includegraphics[width=1.0\linewidth]{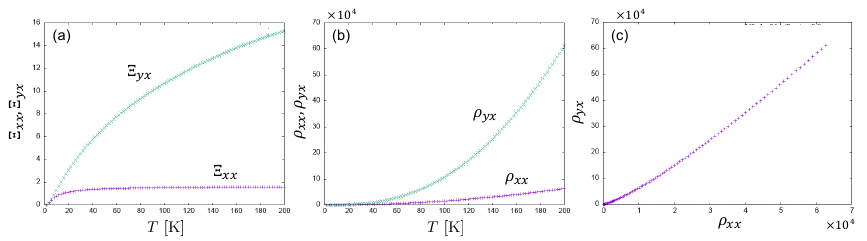}
\caption{\textbf{ Temperature dependence of coefficients $\Xi_{xx}$ and $\Xi_{yx}$.} (a,b) Temperature dependence of $\Xi_{xx}$ and $\Xi_{yx}$ (a), $\bar{\rho}_{xx} \equiv T^2 \Xi_{xx} $ and $\bar{\rho}_{yx} \equiv T^2 \Xi_{yx}$ (b). (c) Scaling between $\bar{\rho}_{xx}$ and $\bar{\rho}_{yx}$. In the calculations, $J S^2 \sim 1200 \, $K\AA$^2$, $q_-  \simeq 0.1 \, {\rm \AA}^{-1}$, and $q_+  \simeq  1.0 \, {\rm \AA}^{-1}$ are used. $q_\pm$ are obtained from the first principles calculations.}
\label{fig:calc}
\end{figure*}

From its form, $H_{\rm so}$ is expected to induce magnon skew scattering. 
The resulting Hall resistivity was calculated in Ref.~\cite{irkhin2021condmat}. 
Here, we recalculate $\sigma_{yx}$ based on the Kubo formula with Feynman diagrams 
of skew-scattering type: see Fig.~\ref{fig:diagram}. 
We also recalculate $\rho_{xx}$ using the force correlation formula\cite{irkhin2021condmat}. 
 The results are given by 
\begin{align}
 \Delta \tilde{\rho}_{yx}^\mathrm{cal}  
&=  - \Xi_{yx} \ \mathrm{sgn}(M_z) \frac{ \lambda a_0^2 J_{pd}^3 S }{8 \hbar^2 e^2 v_{\rm F}^3}  \left( \frac{k_{\rm B}T}{A_\mathrm{ex}} \right)^2, 
\label{eq:rho_xy_summary}\\
  \rho_{xx}^\mathrm{cal}
&=   \Xi_{xx} \frac{ m^2 J_{pd}^2 S}{(2 \pi)^3 \hbar^3 e^2 n^2 a_0^3}  
     \left( \frac{k_{\rm B}T}{A_\mathrm{ex}} \right)^2,  
\label{eq:Irk_xx_summary}
\end{align}
where $M_z$ is the out-of-plane ($z$) component of the magnetization (sgn($x$) is meant to take the sign of $x$), $n$ is the electron density, $v_{\rm F}$ is the Fermi velocity (whose spin dependence is neglected), $A_\mathrm{ex} = J S^2 / a_0^3$ is the exchange stiffness constant, and $\Xi_{xx}$ and $\Xi_{yx}$ are given by
\begin{align}
  \Xi_{xx} &=  \int_{\beta \Omega_-}^{\beta \Omega_+} \frac{xdx}{e^x - 1}   , 
\label{eq:Irk_Psi}
\\
 \Xi_{yx}  &=  \int_{-\infty}^\infty d x \, \left( - \frac{\partial f}{\partial x} \right)  
   \left[ \ln \frac{e^{\beta \Omega_+} - 1}{e^{\beta \Omega_-} - 1} 
          - \ln \frac{e^{\beta \Omega_+} + e^x}{e^{\beta \Omega_-} + e^x}  \right]^2     , 
\label{eq:Xi_s}
\end{align}
with 
$\Omega_\pm  = J S^2 q_\pm^2$, 
$q_\pm  = |k_{{\rm F}\uparrow} \pm k_{{\rm F}\downarrow}|$ 
($k_{{\rm F} \sigma}$: Fermi wave number) and $\beta = (k_{\rm B} T)^{-1}$, and $f$ is the Fermi-Dirac distribution function.
 The temperature dependence of $\Xi_{xx}$ and $\Xi_{yx}$ are shown in Fig.~\ref{fig:calc}(a).  
To see the temperature dependence of the resistivities, we plot $T^2 \Xi_{xx}$ ($\propto \rho_{xx}$) 
and $T^2 \Xi_{yx}$ ($\propto \rho_{yx}$) in Fig.~\ref{fig:calc}(b). 
Even with $\Xi_{xx}$ and $\Xi_{yx}$, they still behave like $\sim T^2$, 
and we see from Fig.~\ref{fig:calc}(c) that $\rho_{yx}$ is practically proportional to $\rho_{xx}$. 

As for the ratio, with $M_z > 0$, 
\begin{align}
 \frac{\Delta \tilde{\rho}_{yx}^\mathrm{cal} }{\rho_{xx}^\mathrm{cal}}  
&= - \frac{\Xi_{yx}}{\Xi_{xx} } \cdot \frac{\pi^3 \hbar \lambda J_{pd} a_0^2 n^2}{{ m^2 v_{\rm F}^3}}
= - \frac{\pi^3}{2} \frac{\Xi_{yx}}{\Xi_{xx} } \cdot \frac{\lambda J_{pd}}{\varepsilon_{\rm F}} \cdot 
      \frac{(a_0^3 n)^2 }{ a_0 k_{\rm F} }.
\label{eq:ratio}
\end{align}
From the experiments, we obtained $a_2 = \Delta \rho_{yx}/\rho_{xx} \simeq - 0.067$. 
To account for this value, we used the material parameters denoted in Table~\ref{table:supps:para} and adjusted $ \lambda $ to obtain $\lambda \sim -0.84$.
\begin{table*}[tb]
	\centering
	\begin{threeparttable}[b]
	\caption{\label{table:supps:para} \textbf{Calculation parameters.} All parameters used in the calculations are presented. }
	\begin{tabular}{ccccc}
		\hline
		parameter & notation & value & unit & source \\
		\hline \hline
		Volume per localized spin & $a_0^3$ & $6.02 \times 10^{-23}$ & cm$^{3}$ & DFT (Sec.~\ref{sec:supps:dft})\\
		Spin magnetic moment & $S$ & $\frac{3}{2}$ &  & Footnote\tnote{1}\\
		Saturation magnetization & $\mu_0 M_\mathrm{s}$ & $400$ & emu cm$^{-3}$ & Fig.~\ref{fig:rhoxx}\\
		Curie temperature & $T_\mathrm{C}$ & $175$ & K & Sec.~\ref{sec:supps:trans}, Fig.~\ref{fig:vsm}\\
		Exchange stiffness constant & $A_\mathrm{ex}$ & $2.3 \times 10^{-7}$ & erg cm$^{-1}$ & Footnote\tnote{2}\\
		Fermi energy & $\varepsilon_\mathrm{F}$ & $0.3$ & eV & DFT (Sec.~\ref{sec:supps:dft})\\
		Carrier density & $n$ & $7.5 \times 10^{20}$ & cm$^{-3}$ & Footnote\tnote{3}\\
		Fermi velocity & $v_\mathrm{F}$ & $3.2 \times 10^{7}$ & cm s$^{-1}$ & Footnote\tnote{3}\\
		$p$-$d$ exchange interaction & $ J_{pd}$ & $- \frac{0.15}{S} = - 0.1$ & eV & DFT (Sec.~\ref{sec:supps:dft})\\
		Spin orbit parameter & $\lambda$ & $- 0.84$ &   & Fit data\tnote{4}\\
		\hline \hline
	\end{tabular}
		\begin{tablenotes}
        			\item[1] Assumed electron configuration of Cr$^{3+}$.
			\item[2] Calculated using the relation $A_\mathrm{ex} \propto M_\mathrm{s}^{\frac{1}{3}} T_\mathrm{C}$ \cite{kuzmin2020epjp}. NiFe is used as a reference ($A_\mathrm{ex}: 1.3 \times 10^{-6}$ erg cm$^{-1}$, $M_\mathrm{s}: 800$ emu cm$^{-3}$, $T_\mathrm{C}: 800$ K). 
        			\item[3] Assumed a free electron dispersion: $k_\mathrm{F} = \left( \frac{2 m \varepsilon_\mathrm{F}}{\hbar^2} \right)^{\frac{1}{2}}$, $n = \frac{k_\mathrm{F}^3}{3 \pi^2}$, $v_\mathrm{F} = \frac{\hbar k_\mathrm{F} }{m}$.
        			\item[4] Adjust $\lambda$ to match experimentally obtained value of $a_2$ with $\frac{\Delta \tilde{\rho}_{yx}^\mathrm{cal} }{ \rho_{xx}^\mathrm{cal} }$.
        		\end{tablenotes}
        	\end{threeparttable}
\end{table*}

\clearpage
\bibliography{references_041425}

\begin{thebibliography}{50}%
\makeatletter
\providecommand \@ifxundefined [1]{%
 \@ifx{#1\undefined}
}%
\providecommand \@ifnum [1]{%
 \ifnum #1\expandafter \@firstoftwo
 \else \expandafter \@secondoftwo
 \fi
}%
\providecommand \@ifx [1]{%
 \ifx #1\expandafter \@firstoftwo
 \else \expandafter \@secondoftwo
 \fi
}%
\providecommand \natexlab [1]{#1}%
\providecommand \enquote  [1]{``#1''}%
\providecommand \bibnamefont  [1]{#1}%
\providecommand \bibfnamefont [1]{#1}%
\providecommand \citenamefont [1]{#1}%
\providecommand \href@noop [0]{\@secondoftwo}%
\providecommand \href [0]{\begingroup \@sanitize@url \@href}%
\providecommand \@href[1]{\@@startlink{#1}\@@href}%
\providecommand \@@href[1]{\endgroup#1\@@endlink}%
\providecommand \@sanitize@url [0]{\catcode `\\12\catcode `\$12\catcode
  `\&12\catcode `\#12\catcode `\^12\catcode `\_12\catcode `\%12\relax}%
\providecommand \@@startlink[1]{}%
\providecommand \@@endlink[0]{}%
\providecommand \url  [0]{\begingroup\@sanitize@url \@url }%
\providecommand \@url [1]{\endgroup\@href {#1}{\urlprefix }}%
\providecommand \urlprefix  [0]{URL }%
\providecommand \Eprint [0]{\href }%
\providecommand \doibase [0]{https://doi.org/}%
\providecommand \selectlanguage [0]{\@gobble}%
\providecommand \bibinfo  [0]{\@secondoftwo}%
\providecommand \bibfield  [0]{\@secondoftwo}%
\providecommand \translation [1]{[#1]}%
\providecommand \BibitemOpen [0]{}%
\providecommand \bibitemStop [0]{}%
\providecommand \bibitemNoStop [0]{.\EOS\space}%
\providecommand \EOS [0]{\spacefactor3000\relax}%
\providecommand \BibitemShut  [1]{\csname bibitem#1\endcsname}%
\let\auto@bib@innerbib\@empty
\bibitem [{\citenamefont {Ipser}\ \emph {et~al.}(1983)\citenamefont {Ipser},
  \citenamefont {Komarek},\ and\ \citenamefont {Klepp}}]{ipser1983jlesscommon}%
  \BibitemOpen
  \bibfield  {author} {\bibinfo {author} {\bibfnamefont {H.}~\bibnamefont
  {Ipser}}, \bibinfo {author} {\bibfnamefont {K.~L.}\ \bibnamefont {Komarek}},\
  and\ \bibinfo {author} {\bibfnamefont {K.~O.}\ \bibnamefont {Klepp}},\
  }\bibfield  {title} {\bibinfo {title} {Transition metal-chalcogen systems
  viii: The cr-te phase diagram},\ }\href@noop {} {\bibfield  {journal}
  {\bibinfo  {journal} {Journal of the Less Common Metals}\ }\textbf {\bibinfo
  {volume} {92}},\ \bibinfo {pages} {265} (\bibinfo {year} {1983})}\BibitemShut
  {NoStop}%
\bibitem [{\citenamefont {McGuire}\ \emph {et~al.}(2017)\citenamefont
  {McGuire}, \citenamefont {Garlea}, \citenamefont {Santosh}, \citenamefont
  {Cooper}, \citenamefont {Yan}, \citenamefont {Cao},\ and\ \citenamefont
  {Sales}}]{mcguire2017prb}%
  \BibitemOpen
  \bibfield  {author} {\bibinfo {author} {\bibfnamefont {M.~A.}\ \bibnamefont
  {McGuire}}, \bibinfo {author} {\bibfnamefont {V.~O.}\ \bibnamefont {Garlea}},
  \bibinfo {author} {\bibfnamefont {K.~C.}\ \bibnamefont {Santosh}}, \bibinfo
  {author} {\bibfnamefont {V.~R.}\ \bibnamefont {Cooper}}, \bibinfo {author}
  {\bibfnamefont {J.~Q.}\ \bibnamefont {Yan}}, \bibinfo {author} {\bibfnamefont
  {H.~B.}\ \bibnamefont {Cao}},\ and\ \bibinfo {author} {\bibfnamefont {B.~C.}\
  \bibnamefont {Sales}},\ }\bibfield  {title} {\bibinfo {title}
  {Antiferromagnetism in the van der waals layered spin-lozenge semiconductor
  crte$_3$},\ }\href@noop {} {\bibfield  {journal} {\bibinfo  {journal} {Phys.
  Rev. B}\ }\textbf {\bibinfo {volume} {95}},\ \bibinfo {pages} {144421}
  (\bibinfo {year} {2017})}\BibitemShut {NoStop}%
\bibitem [{\citenamefont {Fujisawa}\ \emph {et~al.}(2020)\citenamefont
  {Fujisawa}, \citenamefont {Pardo-Almanza}, \citenamefont {Garland},
  \citenamefont {Yamagami}, \citenamefont {Zhu}, \citenamefont {Chen},
  \citenamefont {Araki}, \citenamefont {Takeda}, \citenamefont {Kobayashi},
  \citenamefont {Takeda}, \citenamefont {Hsu}, \citenamefont {Chuang},
  \citenamefont {Laskowski}, \citenamefont {Khoo}, \citenamefont
  {Soumyanarayanan},\ and\ \citenamefont {Okada}}]{fujisawa2020prm}%
  \BibitemOpen
  \bibfield  {author} {\bibinfo {author} {\bibfnamefont {Y.}~\bibnamefont
  {Fujisawa}}, \bibinfo {author} {\bibfnamefont {M.}~\bibnamefont
  {Pardo-Almanza}}, \bibinfo {author} {\bibfnamefont {J.}~\bibnamefont
  {Garland}}, \bibinfo {author} {\bibfnamefont {K.}~\bibnamefont {Yamagami}},
  \bibinfo {author} {\bibfnamefont {X.}~\bibnamefont {Zhu}}, \bibinfo {author}
  {\bibfnamefont {X.}~\bibnamefont {Chen}}, \bibinfo {author} {\bibfnamefont
  {K.}~\bibnamefont {Araki}}, \bibinfo {author} {\bibfnamefont
  {T.}~\bibnamefont {Takeda}}, \bibinfo {author} {\bibfnamefont
  {M.}~\bibnamefont {Kobayashi}}, \bibinfo {author} {\bibfnamefont
  {Y.}~\bibnamefont {Takeda}}, \bibinfo {author} {\bibfnamefont {C.~H.}\
  \bibnamefont {Hsu}}, \bibinfo {author} {\bibfnamefont {F.~C.}\ \bibnamefont
  {Chuang}}, \bibinfo {author} {\bibfnamefont {R.}~\bibnamefont {Laskowski}},
  \bibinfo {author} {\bibfnamefont {K.~H.}\ \bibnamefont {Khoo}}, \bibinfo
  {author} {\bibfnamefont {A.}~\bibnamefont {Soumyanarayanan}},\ and\ \bibinfo
  {author} {\bibfnamefont {Y.}~\bibnamefont {Okada}},\ }\bibfield  {title}
  {\bibinfo {title} {Tailoring magnetism in self-intercalated
  cr$_{1+\delta}$te$_2$ epitaxial films},\ }\href@noop {} {\bibfield  {journal}
  {\bibinfo  {journal} {Phys. Rev. Mater.}\ }\textbf {\bibinfo {volume} {4}},\
  \bibinfo {pages} {114001} (\bibinfo {year} {2020})}\BibitemShut {NoStop}%
\bibitem [{\citenamefont {Zhang}\ \emph
  {et~al.}(2021{\natexlab{a}})\citenamefont {Zhang}, \citenamefont {Lu},
  \citenamefont {Liu}, \citenamefont {Niu}, \citenamefont {Sun}, \citenamefont
  {Cook}, \citenamefont {Vaninger}, \citenamefont {Miceli}, \citenamefont
  {Singh}, \citenamefont {Lian}, \citenamefont {Chang}, \citenamefont {He},
  \citenamefont {Du}, \citenamefont {He}, \citenamefont {Zhang}, \citenamefont
  {Bian},\ and\ \citenamefont {Xu}}]{zhang2021ncomm}%
  \BibitemOpen
  \bibfield  {author} {\bibinfo {author} {\bibfnamefont {X.~Q.}\ \bibnamefont
  {Zhang}}, \bibinfo {author} {\bibfnamefont {Q.~S.}\ \bibnamefont {Lu}},
  \bibinfo {author} {\bibfnamefont {W.~Q.}\ \bibnamefont {Liu}}, \bibinfo
  {author} {\bibfnamefont {W.}~\bibnamefont {Niu}}, \bibinfo {author}
  {\bibfnamefont {J.~B.}\ \bibnamefont {Sun}}, \bibinfo {author} {\bibfnamefont
  {J.}~\bibnamefont {Cook}}, \bibinfo {author} {\bibfnamefont {M.}~\bibnamefont
  {Vaninger}}, \bibinfo {author} {\bibfnamefont {P.~F.}\ \bibnamefont
  {Miceli}}, \bibinfo {author} {\bibfnamefont {D.~J.}\ \bibnamefont {Singh}},
  \bibinfo {author} {\bibfnamefont {S.~W.}\ \bibnamefont {Lian}}, \bibinfo
  {author} {\bibfnamefont {T.~R.}\ \bibnamefont {Chang}}, \bibinfo {author}
  {\bibfnamefont {X.~Q.}\ \bibnamefont {He}}, \bibinfo {author} {\bibfnamefont
  {J.}~\bibnamefont {Du}}, \bibinfo {author} {\bibfnamefont {L.}~\bibnamefont
  {He}}, \bibinfo {author} {\bibfnamefont {R.}~\bibnamefont {Zhang}}, \bibinfo
  {author} {\bibfnamefont {G.}~\bibnamefont {Bian}},\ and\ \bibinfo {author}
  {\bibfnamefont {Y.~B.}\ \bibnamefont {Xu}},\ }\bibfield  {title} {\bibinfo
  {title} {Room-temperature intrinsic ferromagnetism in epitaxial crte$_2$
  ultrathin films},\ }\href@noop {} {\bibfield  {journal} {\bibinfo  {journal}
  {Nat. Commun.}\ }\textbf {\bibinfo {volume} {12}},\ \bibinfo {pages} {2492}
  (\bibinfo {year} {2021}{\natexlab{a}})}\BibitemShut {NoStop}%
\bibitem [{\citenamefont {Huang}\ \emph {et~al.}(2017)\citenamefont {Huang},
  \citenamefont {Clark}, \citenamefont {Navarro-Moratalla}, \citenamefont
  {Klein}, \citenamefont {Cheng}, \citenamefont {Seyler}, \citenamefont
  {Zhong}, \citenamefont {Schmidgall}, \citenamefont {McGuire}, \citenamefont
  {Cobden}, \citenamefont {Yao}, \citenamefont {Xiao}, \citenamefont
  {Jarillo-Herrero},\ and\ \citenamefont {Xu}}]{huang2017nature}%
  \BibitemOpen
  \bibfield  {author} {\bibinfo {author} {\bibfnamefont {B.}~\bibnamefont
  {Huang}}, \bibinfo {author} {\bibfnamefont {G.}~\bibnamefont {Clark}},
  \bibinfo {author} {\bibfnamefont {E.}~\bibnamefont {Navarro-Moratalla}},
  \bibinfo {author} {\bibfnamefont {D.~R.}\ \bibnamefont {Klein}}, \bibinfo
  {author} {\bibfnamefont {R.}~\bibnamefont {Cheng}}, \bibinfo {author}
  {\bibfnamefont {K.~L.}\ \bibnamefont {Seyler}}, \bibinfo {author}
  {\bibfnamefont {D.}~\bibnamefont {Zhong}}, \bibinfo {author} {\bibfnamefont
  {E.}~\bibnamefont {Schmidgall}}, \bibinfo {author} {\bibfnamefont {M.~A.}\
  \bibnamefont {McGuire}}, \bibinfo {author} {\bibfnamefont {D.~H.}\
  \bibnamefont {Cobden}}, \bibinfo {author} {\bibfnamefont {W.}~\bibnamefont
  {Yao}}, \bibinfo {author} {\bibfnamefont {D.}~\bibnamefont {Xiao}}, \bibinfo
  {author} {\bibfnamefont {P.}~\bibnamefont {Jarillo-Herrero}},\ and\ \bibinfo
  {author} {\bibfnamefont {X.}~\bibnamefont {Xu}},\ }\bibfield  {title}
  {\bibinfo {title} {Layer-dependent ferromagnetism in a van der waals crystal
  down to the monolayer limit},\ }\href@noop {} {\bibfield  {journal} {\bibinfo
   {journal} {Nature}\ }\textbf {\bibinfo {volume} {546}},\ \bibinfo {pages}
  {270} (\bibinfo {year} {2017})}\BibitemShut {NoStop}%
\bibitem [{\citenamefont {Gong}\ \emph {et~al.}(2017)\citenamefont {Gong},
  \citenamefont {Li}, \citenamefont {Li}, \citenamefont {Ji}, \citenamefont
  {Stern}, \citenamefont {Xia}, \citenamefont {Cao}, \citenamefont {Bao},
  \citenamefont {Wang}, \citenamefont {Wang}, \citenamefont {Qiu},
  \citenamefont {Cava}, \citenamefont {Louie}, \citenamefont {Xia},\ and\
  \citenamefont {Zhang}}]{gong2017nature}%
  \BibitemOpen
  \bibfield  {author} {\bibinfo {author} {\bibfnamefont {C.}~\bibnamefont
  {Gong}}, \bibinfo {author} {\bibfnamefont {L.}~\bibnamefont {Li}}, \bibinfo
  {author} {\bibfnamefont {Z.}~\bibnamefont {Li}}, \bibinfo {author}
  {\bibfnamefont {H.}~\bibnamefont {Ji}}, \bibinfo {author} {\bibfnamefont
  {A.}~\bibnamefont {Stern}}, \bibinfo {author} {\bibfnamefont
  {Y.}~\bibnamefont {Xia}}, \bibinfo {author} {\bibfnamefont {T.}~\bibnamefont
  {Cao}}, \bibinfo {author} {\bibfnamefont {W.}~\bibnamefont {Bao}}, \bibinfo
  {author} {\bibfnamefont {C.}~\bibnamefont {Wang}}, \bibinfo {author}
  {\bibfnamefont {Y.}~\bibnamefont {Wang}}, \bibinfo {author} {\bibfnamefont
  {Z.~Q.}\ \bibnamefont {Qiu}}, \bibinfo {author} {\bibfnamefont {R.~J.}\
  \bibnamefont {Cava}}, \bibinfo {author} {\bibfnamefont {S.~G.}\ \bibnamefont
  {Louie}}, \bibinfo {author} {\bibfnamefont {J.}~\bibnamefont {Xia}},\ and\
  \bibinfo {author} {\bibfnamefont {X.}~\bibnamefont {Zhang}},\ }\bibfield
  {title} {\bibinfo {title} {Discovery of intrinsic ferromagnetism in
  two-dimensional van der waals crystals},\ }\href@noop {} {\bibfield
  {journal} {\bibinfo  {journal} {Nature}\ }\textbf {\bibinfo {volume} {546}},\
  \bibinfo {pages} {265} (\bibinfo {year} {2017})}\BibitemShut {NoStop}%
\bibitem [{\citenamefont {Gibertini}\ \emph {et~al.}(2019)\citenamefont
  {Gibertini}, \citenamefont {Koperski}, \citenamefont {Morpurgo},\ and\
  \citenamefont {Novoselov}}]{gibertini2019nnano}%
  \BibitemOpen
  \bibfield  {author} {\bibinfo {author} {\bibfnamefont {M.}~\bibnamefont
  {Gibertini}}, \bibinfo {author} {\bibfnamefont {M.}~\bibnamefont {Koperski}},
  \bibinfo {author} {\bibfnamefont {A.~F.}\ \bibnamefont {Morpurgo}},\ and\
  \bibinfo {author} {\bibfnamefont {K.~S.}\ \bibnamefont {Novoselov}},\
  }\bibfield  {title} {\bibinfo {title} {Magnetic 2d materials and
  heterostructures},\ }\href@noop {} {\bibfield  {journal} {\bibinfo  {journal}
  {Nat. Nanotechnol.}\ }\textbf {\bibinfo {volume} {14}},\ \bibinfo {pages}
  {408} (\bibinfo {year} {2019})}\BibitemShut {NoStop}%
\bibitem [{\citenamefont {Zhong}\ \emph {et~al.}(2022)\citenamefont {Zhong},
  \citenamefont {Wang}, \citenamefont {Liu}, \citenamefont {Zhao},
  \citenamefont {Xu}, \citenamefont {Zhou}, \citenamefont {Han}, \citenamefont
  {Gan},\ and\ \citenamefont {Zhai}}]{zhong2022nanores}%
  \BibitemOpen
  \bibfield  {author} {\bibinfo {author} {\bibfnamefont {J.~C.}\ \bibnamefont
  {Zhong}}, \bibinfo {author} {\bibfnamefont {M.~S.}\ \bibnamefont {Wang}},
  \bibinfo {author} {\bibfnamefont {T.}~\bibnamefont {Liu}}, \bibinfo {author}
  {\bibfnamefont {Y.~H.}\ \bibnamefont {Zhao}}, \bibinfo {author}
  {\bibfnamefont {X.}~\bibnamefont {Xu}}, \bibinfo {author} {\bibfnamefont
  {S.~S.}\ \bibnamefont {Zhou}}, \bibinfo {author} {\bibfnamefont {J.~B.}\
  \bibnamefont {Han}}, \bibinfo {author} {\bibfnamefont {L.}~\bibnamefont
  {Gan}},\ and\ \bibinfo {author} {\bibfnamefont {T.~Y.}\ \bibnamefont
  {Zhai}},\ }\bibfield  {title} {\bibinfo {title} {Strain-sensitive
  ferromagnetic two-dimensional cr$_2$te$_3$},\ }\href@noop {} {\bibfield
  {journal} {\bibinfo  {journal} {Nano Research}\ }\textbf {\bibinfo {volume}
  {15}},\ \bibinfo {pages} {1254} (\bibinfo {year} {2022})}\BibitemShut
  {NoStop}%
\bibitem [{\citenamefont {Li}\ \emph {et~al.}(2019)\citenamefont {Li},
  \citenamefont {Wang}, \citenamefont {Chen}, \citenamefont {Yu}, \citenamefont
  {Zhou}, \citenamefont {Qiu}, \citenamefont {He}, \citenamefont {Ye},
  \citenamefont {Sou},\ and\ \citenamefont {Wang}}]{li2019acsapplnanomater}%
  \BibitemOpen
  \bibfield  {author} {\bibinfo {author} {\bibfnamefont {H.~X.}\ \bibnamefont
  {Li}}, \bibinfo {author} {\bibfnamefont {L.~J.}\ \bibnamefont {Wang}},
  \bibinfo {author} {\bibfnamefont {J.~S.}\ \bibnamefont {Chen}}, \bibinfo
  {author} {\bibfnamefont {T.}~\bibnamefont {Yu}}, \bibinfo {author}
  {\bibfnamefont {L.}~\bibnamefont {Zhou}}, \bibinfo {author} {\bibfnamefont
  {Y.}~\bibnamefont {Qiu}}, \bibinfo {author} {\bibfnamefont {H.~T.}\
  \bibnamefont {He}}, \bibinfo {author} {\bibfnamefont {F.}~\bibnamefont {Ye}},
  \bibinfo {author} {\bibfnamefont {T.~K.}\ \bibnamefont {Sou}},\ and\ \bibinfo
  {author} {\bibfnamefont {G.}~\bibnamefont {Wang}},\ }\bibfield  {title}
  {\bibinfo {title} {Molecular beam epitaxy grown cr$_2$te$_3$ thin films with
  tunable curie temperatures for spintronic devices},\ }\href@noop {}
  {\bibfield  {journal} {\bibinfo  {journal} {Acs Applied Nano Materials}\
  }\textbf {\bibinfo {volume} {2}},\ \bibinfo {pages} {6809} (\bibinfo {year}
  {2019})}\BibitemShut {NoStop}%
\bibitem [{\citenamefont {Wen}\ \emph {et~al.}(2020)\citenamefont {Wen},
  \citenamefont {Liu}, \citenamefont {Zhang}, \citenamefont {Xia},
  \citenamefont {Zhai}, \citenamefont {Zhang}, \citenamefont {Zhai},
  \citenamefont {Shen}, \citenamefont {He}, \citenamefont {Cheng},
  \citenamefont {Yin}, \citenamefont {Yao}, \citenamefont {Sendeku},
  \citenamefont {Wang}, \citenamefont {Ye}, \citenamefont {Liu}, \citenamefont
  {Jiang}, \citenamefont {Shan}, \citenamefont {Long},\ and\ \citenamefont
  {He}}]{wen2020nanolett}%
  \BibitemOpen
  \bibfield  {author} {\bibinfo {author} {\bibfnamefont {Y.}~\bibnamefont
  {Wen}}, \bibinfo {author} {\bibfnamefont {Z.~H.}\ \bibnamefont {Liu}},
  \bibinfo {author} {\bibfnamefont {Y.}~\bibnamefont {Zhang}}, \bibinfo
  {author} {\bibfnamefont {C.~X.}\ \bibnamefont {Xia}}, \bibinfo {author}
  {\bibfnamefont {B.~X.}\ \bibnamefont {Zhai}}, \bibinfo {author}
  {\bibfnamefont {X.~H.}\ \bibnamefont {Zhang}}, \bibinfo {author}
  {\bibfnamefont {G.~H.}\ \bibnamefont {Zhai}}, \bibinfo {author}
  {\bibfnamefont {C.}~\bibnamefont {Shen}}, \bibinfo {author} {\bibfnamefont
  {P.}~\bibnamefont {He}}, \bibinfo {author} {\bibfnamefont {R.~Q.}\
  \bibnamefont {Cheng}}, \bibinfo {author} {\bibfnamefont {L.}~\bibnamefont
  {Yin}}, \bibinfo {author} {\bibfnamefont {Y.~Y.}\ \bibnamefont {Yao}},
  \bibinfo {author} {\bibfnamefont {M.~G.}\ \bibnamefont {Sendeku}}, \bibinfo
  {author} {\bibfnamefont {Z.~X.}\ \bibnamefont {Wang}}, \bibinfo {author}
  {\bibfnamefont {X.~B.}\ \bibnamefont {Ye}}, \bibinfo {author} {\bibfnamefont
  {C.~S.}\ \bibnamefont {Liu}}, \bibinfo {author} {\bibfnamefont
  {C.}~\bibnamefont {Jiang}}, \bibinfo {author} {\bibfnamefont {C.~X.}\
  \bibnamefont {Shan}}, \bibinfo {author} {\bibfnamefont {Y.~W.}\ \bibnamefont
  {Long}},\ and\ \bibinfo {author} {\bibfnamefont {J.}~\bibnamefont {He}},\
  }\bibfield  {title} {\bibinfo {title} {Tunable room-temperature
  ferromagnetism in two-dimensional cr$_2$te$_3$},\ }\href@noop {} {\bibfield
  {journal} {\bibinfo  {journal} {Nano Lett.}\ }\textbf {\bibinfo {volume}
  {20}},\ \bibinfo {pages} {3130} (\bibinfo {year} {2020})}\BibitemShut
  {NoStop}%
\bibitem [{\citenamefont {Chua}\ \emph {et~al.}(2021)\citenamefont {Chua},
  \citenamefont {Zhou}, \citenamefont {Yu}, \citenamefont {Yu}, \citenamefont
  {Gou}, \citenamefont {Zhu}, \citenamefont {Zhang}, \citenamefont {Liu},
  \citenamefont {Breese}, \citenamefont {Chen}, \citenamefont {Loh},
  \citenamefont {Feng}, \citenamefont {Yang}, \citenamefont {Huang},\ and\
  \citenamefont {Wee}}]{chua2021advmater}%
  \BibitemOpen
  \bibfield  {author} {\bibinfo {author} {\bibfnamefont {R.}~\bibnamefont
  {Chua}}, \bibinfo {author} {\bibfnamefont {J.}~\bibnamefont {Zhou}}, \bibinfo
  {author} {\bibfnamefont {X.~J.}\ \bibnamefont {Yu}}, \bibinfo {author}
  {\bibfnamefont {W.}~\bibnamefont {Yu}}, \bibinfo {author} {\bibfnamefont
  {J.}~\bibnamefont {Gou}}, \bibinfo {author} {\bibfnamefont {R.}~\bibnamefont
  {Zhu}}, \bibinfo {author} {\bibfnamefont {L.}~\bibnamefont {Zhang}}, \bibinfo
  {author} {\bibfnamefont {M.~Z.}\ \bibnamefont {Liu}}, \bibinfo {author}
  {\bibfnamefont {M.~B.~H.}\ \bibnamefont {Breese}}, \bibinfo {author}
  {\bibfnamefont {W.}~\bibnamefont {Chen}}, \bibinfo {author} {\bibfnamefont
  {K.~P.}\ \bibnamefont {Loh}}, \bibinfo {author} {\bibfnamefont {Y.~P.}\
  \bibnamefont {Feng}}, \bibinfo {author} {\bibfnamefont {M.}~\bibnamefont
  {Yang}}, \bibinfo {author} {\bibfnamefont {Y.~L.}\ \bibnamefont {Huang}},\
  and\ \bibinfo {author} {\bibfnamefont {A.~T.~S.}\ \bibnamefont {Wee}},\
  }\bibfield  {title} {\bibinfo {title} {Room temperature ferromagnetism of
  monolayer chromium telluride with perpendicular magnetic anisotropy},\
  }\href@noop {} {\bibfield  {journal} {\bibinfo  {journal} {Adv. Mater.}\
  }\textbf {\bibinfo {volume} {33}},\ \bibinfo {pages} {2103360} (\bibinfo
  {year} {2021})}\BibitemShut {NoStop}%
\bibitem [{\citenamefont {Coughlin}\ \emph {et~al.}(2020)\citenamefont
  {Coughlin}, \citenamefont {Xie}, \citenamefont {Yao}, \citenamefont {Zhan},
  \citenamefont {Chen}, \citenamefont {Hewa-Walpitage}, \citenamefont {Zhang},
  \citenamefont {Guo}, \citenamefont {Zhou}, \citenamefont {Lou}, \citenamefont
  {Wang}, \citenamefont {Li}, \citenamefont {Fertig},\ and\ \citenamefont
  {Zhang}}]{coughlin2020acsnano}%
  \BibitemOpen
  \bibfield  {author} {\bibinfo {author} {\bibfnamefont {A.~L.}\ \bibnamefont
  {Coughlin}}, \bibinfo {author} {\bibfnamefont {D.~Y.}\ \bibnamefont {Xie}},
  \bibinfo {author} {\bibfnamefont {Y.}~\bibnamefont {Yao}}, \bibinfo {author}
  {\bibfnamefont {X.}~\bibnamefont {Zhan}}, \bibinfo {author} {\bibfnamefont
  {Q.}~\bibnamefont {Chen}}, \bibinfo {author} {\bibfnamefont {H.}~\bibnamefont
  {Hewa-Walpitage}}, \bibinfo {author} {\bibfnamefont {X.~H.}\ \bibnamefont
  {Zhang}}, \bibinfo {author} {\bibfnamefont {H.}~\bibnamefont {Guo}}, \bibinfo
  {author} {\bibfnamefont {H.~D.}\ \bibnamefont {Zhou}}, \bibinfo {author}
  {\bibfnamefont {J.}~\bibnamefont {Lou}}, \bibinfo {author} {\bibfnamefont
  {J.}~\bibnamefont {Wang}}, \bibinfo {author} {\bibfnamefont {Y.~S.}\
  \bibnamefont {Li}}, \bibinfo {author} {\bibfnamefont {H.~A.}\ \bibnamefont
  {Fertig}},\ and\ \bibinfo {author} {\bibfnamefont {S.~X.}\ \bibnamefont
  {Zhang}},\ }\bibfield  {title} {\bibinfo {title} {Near degeneracy of magnetic
  phases in two-dimensional chromium telluride with enhanced perpendicular
  magnetic anisotropy},\ }\href@noop {} {\bibfield  {journal} {\bibinfo
  {journal} {ACS Nano}\ }\textbf {\bibinfo {volume} {14}},\ \bibinfo {pages}
  {15256} (\bibinfo {year} {2020})}\BibitemShut {NoStop}%
\bibitem [{\citenamefont {Bian}\ \emph {et~al.}(2021)\citenamefont {Bian},
  \citenamefont {Kamenskii}, \citenamefont {Han}, \citenamefont {Li},
  \citenamefont {Wei}, \citenamefont {Tian}, \citenamefont {Eason},
  \citenamefont {Sun}, \citenamefont {He}, \citenamefont {Hui}, \citenamefont
  {Yao}, \citenamefont {Sabirianov}, \citenamefont {Bird}, \citenamefont
  {Yang}, \citenamefont {Miao}, \citenamefont {Lin}, \citenamefont {Crooker},
  \citenamefont {Hou},\ and\ \citenamefont {Zeng}}]{bian2021mrl}%
  \BibitemOpen
  \bibfield  {author} {\bibinfo {author} {\bibfnamefont {M.~Y.}\ \bibnamefont
  {Bian}}, \bibinfo {author} {\bibfnamefont {A.~N.}\ \bibnamefont {Kamenskii}},
  \bibinfo {author} {\bibfnamefont {M.~J.}\ \bibnamefont {Han}}, \bibinfo
  {author} {\bibfnamefont {W.~J.}\ \bibnamefont {Li}}, \bibinfo {author}
  {\bibfnamefont {S.~C.}\ \bibnamefont {Wei}}, \bibinfo {author} {\bibfnamefont
  {X.~Z.}\ \bibnamefont {Tian}}, \bibinfo {author} {\bibfnamefont {D.~B.}\
  \bibnamefont {Eason}}, \bibinfo {author} {\bibfnamefont {F.}~\bibnamefont
  {Sun}}, \bibinfo {author} {\bibfnamefont {K.~K.}\ \bibnamefont {He}},
  \bibinfo {author} {\bibfnamefont {H.~L.}\ \bibnamefont {Hui}}, \bibinfo
  {author} {\bibfnamefont {F.}~\bibnamefont {Yao}}, \bibinfo {author}
  {\bibfnamefont {R.}~\bibnamefont {Sabirianov}}, \bibinfo {author}
  {\bibfnamefont {J.~P.}\ \bibnamefont {Bird}}, \bibinfo {author}
  {\bibfnamefont {C.~L.}\ \bibnamefont {Yang}}, \bibinfo {author}
  {\bibfnamefont {J.~W.}\ \bibnamefont {Miao}}, \bibinfo {author}
  {\bibfnamefont {J.~H.}\ \bibnamefont {Lin}}, \bibinfo {author} {\bibfnamefont
  {S.~A.}\ \bibnamefont {Crooker}}, \bibinfo {author} {\bibfnamefont {Y.~L.}\
  \bibnamefont {Hou}},\ and\ \bibinfo {author} {\bibfnamefont {H.}~\bibnamefont
  {Zeng}},\ }\bibfield  {title} {\bibinfo {title} {Covalent 2d cr2te3
  ferromagnet},\ }\href@noop {} {\bibfield  {journal} {\bibinfo  {journal}
  {Mater. Res. Lett.}\ }\textbf {\bibinfo {volume} {9}},\ \bibinfo {pages}
  {205} (\bibinfo {year} {2021})}\BibitemShut {NoStop}%
\bibitem [{\citenamefont {Liu}\ and\ \citenamefont
  {Petrovic}(2018)}]{liu2018prb}%
  \BibitemOpen
  \bibfield  {author} {\bibinfo {author} {\bibfnamefont {Y.}~\bibnamefont
  {Liu}}\ and\ \bibinfo {author} {\bibfnamefont {C.}~\bibnamefont {Petrovic}},\
  }\bibfield  {title} {\bibinfo {title} {Anomalous hall effect in the trigonal
  cr$_5$te$_8$ single crystal},\ }\href@noop {} {\bibfield  {journal} {\bibinfo
   {journal} {Phys. Rev. B}\ }\textbf {\bibinfo {volume} {98}},\ \bibinfo
  {pages} {195122} (\bibinfo {year} {2018})}\BibitemShut {NoStop}%
\bibitem [{\citenamefont {Jiang}\ \emph {et~al.}(2020)\citenamefont {Jiang},
  \citenamefont {Luo}, \citenamefont {Yan}, \citenamefont {Gao}, \citenamefont
  {Wang}, \citenamefont {Zhao}, \citenamefont {Sun}, \citenamefont {Si},
  \citenamefont {Lu}, \citenamefont {Tong}, \citenamefont {Zhu}, \citenamefont
  {Song},\ and\ \citenamefont {Sun}}]{jiang2020prb}%
  \BibitemOpen
  \bibfield  {author} {\bibinfo {author} {\bibfnamefont {Z.~Z.}\ \bibnamefont
  {Jiang}}, \bibinfo {author} {\bibfnamefont {X.}~\bibnamefont {Luo}}, \bibinfo
  {author} {\bibfnamefont {J.}~\bibnamefont {Yan}}, \bibinfo {author}
  {\bibfnamefont {J.~J.}\ \bibnamefont {Gao}}, \bibinfo {author} {\bibfnamefont
  {W.}~\bibnamefont {Wang}}, \bibinfo {author} {\bibfnamefont {G.~C.}\
  \bibnamefont {Zhao}}, \bibinfo {author} {\bibfnamefont {Y.}~\bibnamefont
  {Sun}}, \bibinfo {author} {\bibfnamefont {J.~G.}\ \bibnamefont {Si}},
  \bibinfo {author} {\bibfnamefont {W.~J.}\ \bibnamefont {Lu}}, \bibinfo
  {author} {\bibfnamefont {P.}~\bibnamefont {Tong}}, \bibinfo {author}
  {\bibfnamefont {X.~B.}\ \bibnamefont {Zhu}}, \bibinfo {author} {\bibfnamefont
  {W.~H.}\ \bibnamefont {Song}},\ and\ \bibinfo {author} {\bibfnamefont
  {Y.~P.}\ \bibnamefont {Sun}},\ }\bibfield  {title} {\bibinfo {title}
  {Magnetic anisotropy and anomalous hall effect in monoclinic single crystal
  cr$_5$te$_8$},\ }\href@noop {} {\bibfield  {journal} {\bibinfo  {journal}
  {Phys. Rev. B}\ }\textbf {\bibinfo {volume} {102}},\ \bibinfo {pages}
  {144433} (\bibinfo {year} {2020})}\BibitemShut {NoStop}%
\bibitem [{\citenamefont {Huang}\ \emph
  {et~al.}(2021{\natexlab{a}})\citenamefont {Huang}, \citenamefont {Wang},
  \citenamefont {Wang}, \citenamefont {Liu}, \citenamefont {Xiang},
  \citenamefont {Feng}, \citenamefont {Wang}, \citenamefont {Zhang},
  \citenamefont {Wen}, \citenamefont {Xu}, \citenamefont {Yu}, \citenamefont
  {Lu}, \citenamefont {Zhao}, \citenamefont {Yang}, \citenamefont {Hou},\ and\
  \citenamefont {Xiang}}]{huang2021acsnano}%
  \BibitemOpen
  \bibfield  {author} {\bibinfo {author} {\bibfnamefont {M.}~\bibnamefont
  {Huang}}, \bibinfo {author} {\bibfnamefont {S.~S.}\ \bibnamefont {Wang}},
  \bibinfo {author} {\bibfnamefont {Z.~H.}\ \bibnamefont {Wang}}, \bibinfo
  {author} {\bibfnamefont {P.}~\bibnamefont {Liu}}, \bibinfo {author}
  {\bibfnamefont {J.~X.}\ \bibnamefont {Xiang}}, \bibinfo {author}
  {\bibfnamefont {C.}~\bibnamefont {Feng}}, \bibinfo {author} {\bibfnamefont
  {X.~Q.}\ \bibnamefont {Wang}}, \bibinfo {author} {\bibfnamefont {Z.~M.}\
  \bibnamefont {Zhang}}, \bibinfo {author} {\bibfnamefont {Z.~C.}\ \bibnamefont
  {Wen}}, \bibinfo {author} {\bibfnamefont {H.~J.}\ \bibnamefont {Xu}},
  \bibinfo {author} {\bibfnamefont {G.~Q.}\ \bibnamefont {Yu}}, \bibinfo
  {author} {\bibfnamefont {Y.~L.}\ \bibnamefont {Lu}}, \bibinfo {author}
  {\bibfnamefont {W.~S.}\ \bibnamefont {Zhao}}, \bibinfo {author}
  {\bibfnamefont {S.~Y.~A.}\ \bibnamefont {Yang}}, \bibinfo {author}
  {\bibfnamefont {D.~Z.}\ \bibnamefont {Hou}},\ and\ \bibinfo {author}
  {\bibfnamefont {B.}~\bibnamefont {Xiang}},\ }\bibfield  {title} {\bibinfo
  {title} {Colossal anomalous hall effect in ferromagnetic van der waals
  crte$_2$},\ }\href@noop {} {\bibfield  {journal} {\bibinfo  {journal} {Acs
  Nano}\ }\textbf {\bibinfo {volume} {15}},\ \bibinfo {pages} {9759} (\bibinfo
  {year} {2021}{\natexlab{a}})}\BibitemShut {NoStop}%
\bibitem [{\citenamefont {Sun}\ \emph {et~al.}(2021)\citenamefont {Sun},
  \citenamefont {Yan}, \citenamefont {Ning}, \citenamefont {Zhang},
  \citenamefont {Zhao}, \citenamefont {Gao}, \citenamefont {Kanagaraj},
  \citenamefont {Zhang}, \citenamefont {Li}, \citenamefont {Lu}, \citenamefont
  {Yan}, \citenamefont {Li}, \citenamefont {Xu},\ and\ \citenamefont
  {He}}]{sun2021aip}%
  \BibitemOpen
  \bibfield  {author} {\bibinfo {author} {\bibfnamefont {Y.~Z.}\ \bibnamefont
  {Sun}}, \bibinfo {author} {\bibfnamefont {P.~F.}\ \bibnamefont {Yan}},
  \bibinfo {author} {\bibfnamefont {J.~A.}\ \bibnamefont {Ning}}, \bibinfo
  {author} {\bibfnamefont {X.~Q.}\ \bibnamefont {Zhang}}, \bibinfo {author}
  {\bibfnamefont {Y.~F.}\ \bibnamefont {Zhao}}, \bibinfo {author}
  {\bibfnamefont {Q.~W.}\ \bibnamefont {Gao}}, \bibinfo {author} {\bibfnamefont
  {M.}~\bibnamefont {Kanagaraj}}, \bibinfo {author} {\bibfnamefont {K.~P.}\
  \bibnamefont {Zhang}}, \bibinfo {author} {\bibfnamefont {J.~J.}\ \bibnamefont
  {Li}}, \bibinfo {author} {\bibfnamefont {X.~Y.}\ \bibnamefont {Lu}}, \bibinfo
  {author} {\bibfnamefont {Y.}~\bibnamefont {Yan}}, \bibinfo {author}
  {\bibfnamefont {Y.}~\bibnamefont {Li}}, \bibinfo {author} {\bibfnamefont
  {Y.~B.}\ \bibnamefont {Xu}},\ and\ \bibinfo {author} {\bibfnamefont
  {L.}~\bibnamefont {He}},\ }\bibfield  {title} {\bibinfo {title}
  {Ferromagnetism in two-dimensional crte$_2$ epitaxial films down to a few
  atomic layers},\ }\href@noop {} {\bibfield  {journal} {\bibinfo  {journal}
  {Aip Advances}\ }\textbf {\bibinfo {volume} {11}},\ \bibinfo {pages} {035138}
  (\bibinfo {year} {2021})}\BibitemShut {NoStop}%
\bibitem [{\citenamefont {Ou}\ \emph {et~al.}(2022)\citenamefont {Ou},
  \citenamefont {Yanez}, \citenamefont {Xiao}, \citenamefont {Stanley},
  \citenamefont {Ghosh}, \citenamefont {Zheng}, \citenamefont {Jiang},
  \citenamefont {Huang}, \citenamefont {Pillsbury}, \citenamefont
  {Richardella}, \citenamefont {Liu}, \citenamefont {Low}, \citenamefont
  {Crespi}, \citenamefont {Mkhoyan},\ and\ \citenamefont
  {Samarth}}]{ou2022ncomm}%
  \BibitemOpen
  \bibfield  {author} {\bibinfo {author} {\bibfnamefont {Y.~X.}\ \bibnamefont
  {Ou}}, \bibinfo {author} {\bibfnamefont {W.}~\bibnamefont {Yanez}}, \bibinfo
  {author} {\bibfnamefont {R.}~\bibnamefont {Xiao}}, \bibinfo {author}
  {\bibfnamefont {M.}~\bibnamefont {Stanley}}, \bibinfo {author} {\bibfnamefont
  {S.}~\bibnamefont {Ghosh}}, \bibinfo {author} {\bibfnamefont {B.~Y.}\
  \bibnamefont {Zheng}}, \bibinfo {author} {\bibfnamefont {W.}~\bibnamefont
  {Jiang}}, \bibinfo {author} {\bibfnamefont {Y.~S.}\ \bibnamefont {Huang}},
  \bibinfo {author} {\bibfnamefont {T.}~\bibnamefont {Pillsbury}}, \bibinfo
  {author} {\bibfnamefont {A.}~\bibnamefont {Richardella}}, \bibinfo {author}
  {\bibfnamefont {C.~X.}\ \bibnamefont {Liu}}, \bibinfo {author} {\bibfnamefont
  {T.}~\bibnamefont {Low}}, \bibinfo {author} {\bibfnamefont {V.~H.}\
  \bibnamefont {Crespi}}, \bibinfo {author} {\bibfnamefont {K.~A.}\
  \bibnamefont {Mkhoyan}},\ and\ \bibinfo {author} {\bibfnamefont
  {N.}~\bibnamefont {Samarth}},\ }\bibfield  {title} {\bibinfo {title}
  {Zrte$_2$/crte$_2$: an epitaxial van der waals platform for spintronics},\
  }\href@noop {} {\bibfield  {journal} {\bibinfo  {journal} {Nat. Commun.}\
  }\textbf {\bibinfo {volume} {13}},\ \bibinfo {pages} {2972} (\bibinfo {year}
  {2022})}\BibitemShut {NoStop}%
\bibitem [{\citenamefont {Cho}\ \emph {et~al.}(2023)\citenamefont {Cho},
  \citenamefont {Lee}, \citenamefont {Lee}, \citenamefont {Kim}, \citenamefont
  {Khim}, \citenamefont {Park}, \citenamefont {Jo}, \citenamefont {Choi},
  \citenamefont {Han}, \citenamefont {Chang},\ and\ \citenamefont
  {Lee}}]{cho2023nanoconv}%
  \BibitemOpen
  \bibfield  {author} {\bibinfo {author} {\bibfnamefont {S.~W.}\ \bibnamefont
  {Cho}}, \bibinfo {author} {\bibfnamefont {I.~H.}\ \bibnamefont {Lee}},
  \bibinfo {author} {\bibfnamefont {Y.~W.}\ \bibnamefont {Lee}}, \bibinfo
  {author} {\bibfnamefont {S.}~\bibnamefont {Kim}}, \bibinfo {author}
  {\bibfnamefont {Y.~G.}\ \bibnamefont {Khim}}, \bibinfo {author}
  {\bibfnamefont {S.~Y.}\ \bibnamefont {Park}}, \bibinfo {author}
  {\bibfnamefont {Y.}~\bibnamefont {Jo}}, \bibinfo {author} {\bibfnamefont
  {J.}~\bibnamefont {Choi}}, \bibinfo {author} {\bibfnamefont {S.~W.}\
  \bibnamefont {Han}}, \bibinfo {author} {\bibfnamefont {Y.~J.}\ \bibnamefont
  {Chang}},\ and\ \bibinfo {author} {\bibfnamefont {S.~Y.}\ \bibnamefont
  {Lee}},\ }\bibfield  {title} {\bibinfo {title} {Investigation of the
  mechanism of the anomalous hall effects in cr$_2$te$_3$/(bisb)$_2$(tese)$_3$
  heterostructure},\ }\href@noop {} {\bibfield  {journal} {\bibinfo  {journal}
  {Nano Convergence}\ }\textbf {\bibinfo {volume} {10}},\ \bibinfo {pages} {2}
  (\bibinfo {year} {2023})}\BibitemShut {NoStop}%
\bibitem [{\citenamefont {Chi}\ \emph {et~al.}(2023)\citenamefont {Chi},
  \citenamefont {Ou}, \citenamefont {Eldred}, \citenamefont {Gao},
  \citenamefont {Kwon}, \citenamefont {Murray}, \citenamefont {Dreyer},
  \citenamefont {Butera}, \citenamefont {Foucher}, \citenamefont {Ambaye},
  \citenamefont {Keum}, \citenamefont {Greenberg}, \citenamefont {Liu},
  \citenamefont {Neupane}, \citenamefont {de~Coster}, \citenamefont {Vail},
  \citenamefont {Taylor}, \citenamefont {Folkes}, \citenamefont {Rong},
  \citenamefont {Yin}, \citenamefont {Lake}, \citenamefont {Ross},
  \citenamefont {Lauter}, \citenamefont {Heiman},\ and\ \citenamefont
  {Moodera}}]{chi2023ncomm}%
  \BibitemOpen
  \bibfield  {author} {\bibinfo {author} {\bibfnamefont {H.}~\bibnamefont
  {Chi}}, \bibinfo {author} {\bibfnamefont {Y.~B.}\ \bibnamefont {Ou}},
  \bibinfo {author} {\bibfnamefont {T.~B.}\ \bibnamefont {Eldred}}, \bibinfo
  {author} {\bibfnamefont {W.~P.}\ \bibnamefont {Gao}}, \bibinfo {author}
  {\bibfnamefont {S.}~\bibnamefont {Kwon}}, \bibinfo {author} {\bibfnamefont
  {J.}~\bibnamefont {Murray}}, \bibinfo {author} {\bibfnamefont
  {M.}~\bibnamefont {Dreyer}}, \bibinfo {author} {\bibfnamefont {R.~E.}\
  \bibnamefont {Butera}}, \bibinfo {author} {\bibfnamefont {A.~C.}\
  \bibnamefont {Foucher}}, \bibinfo {author} {\bibfnamefont {H.}~\bibnamefont
  {Ambaye}}, \bibinfo {author} {\bibfnamefont {J.}~\bibnamefont {Keum}},
  \bibinfo {author} {\bibfnamefont {A.~T.}\ \bibnamefont {Greenberg}}, \bibinfo
  {author} {\bibfnamefont {Y.~H.}\ \bibnamefont {Liu}}, \bibinfo {author}
  {\bibfnamefont {M.~R.}\ \bibnamefont {Neupane}}, \bibinfo {author}
  {\bibfnamefont {G.~J.}\ \bibnamefont {de~Coster}}, \bibinfo {author}
  {\bibfnamefont {O.~A.}\ \bibnamefont {Vail}}, \bibinfo {author}
  {\bibfnamefont {P.~J.}\ \bibnamefont {Taylor}}, \bibinfo {author}
  {\bibfnamefont {P.~A.}\ \bibnamefont {Folkes}}, \bibinfo {author}
  {\bibfnamefont {C.~R.}\ \bibnamefont {Rong}}, \bibinfo {author}
  {\bibfnamefont {G.}~\bibnamefont {Yin}}, \bibinfo {author} {\bibfnamefont
  {R.~K.}\ \bibnamefont {Lake}}, \bibinfo {author} {\bibfnamefont {F.~M.}\
  \bibnamefont {Ross}}, \bibinfo {author} {\bibfnamefont {V.}~\bibnamefont
  {Lauter}}, \bibinfo {author} {\bibfnamefont {D.}~\bibnamefont {Heiman}},\
  and\ \bibinfo {author} {\bibfnamefont {J.~S.}\ \bibnamefont {Moodera}},\
  }\bibfield  {title} {\bibinfo {title} {Strain-tunable berry curvature in
  quasi-two-dimensional chromium telluride},\ }\href@noop {} {\bibfield
  {journal} {\bibinfo  {journal} {Nat. Commun.}\ }\textbf {\bibinfo {volume}
  {14}},\ \bibinfo {pages} {3222} (\bibinfo {year} {2023})}\BibitemShut
  {NoStop}%
\bibitem [{\citenamefont {He}\ \emph {et~al.}(2024)\citenamefont {He},
  \citenamefont {Bian}, \citenamefont {Seddon}, \citenamefont {Jagadish},
  \citenamefont {Mucchietto}, \citenamefont {Ren}, \citenamefont {Kirstein},
  \citenamefont {Asadi}, \citenamefont {Bai}, \citenamefont {Yao},
  \citenamefont {Pan}, \citenamefont {Yu}, \citenamefont {Milde}, \citenamefont
  {Huai}, \citenamefont {Hui}, \citenamefont {Zang}, \citenamefont
  {Sabirianov}, \citenamefont {Cheng}, \citenamefont {Miao}, \citenamefont
  {Xing}, \citenamefont {Shao}, \citenamefont {Crooker}, \citenamefont {Eng},
  \citenamefont {Hou}, \citenamefont {Bird},\ and\ \citenamefont
  {Zeng}}]{he2024advsci}%
  \BibitemOpen
  \bibfield  {author} {\bibinfo {author} {\bibfnamefont {K.~K.}\ \bibnamefont
  {He}}, \bibinfo {author} {\bibfnamefont {M.~Y.}\ \bibnamefont {Bian}},
  \bibinfo {author} {\bibfnamefont {S.~D.}\ \bibnamefont {Seddon}}, \bibinfo
  {author} {\bibfnamefont {K.}~\bibnamefont {Jagadish}}, \bibinfo {author}
  {\bibfnamefont {A.}~\bibnamefont {Mucchietto}}, \bibinfo {author}
  {\bibfnamefont {H.}~\bibnamefont {Ren}}, \bibinfo {author} {\bibfnamefont
  {E.}~\bibnamefont {Kirstein}}, \bibinfo {author} {\bibfnamefont
  {R.}~\bibnamefont {Asadi}}, \bibinfo {author} {\bibfnamefont
  {J.}~\bibnamefont {Bai}}, \bibinfo {author} {\bibfnamefont {C.}~\bibnamefont
  {Yao}}, \bibinfo {author} {\bibfnamefont {S.}~\bibnamefont {Pan}}, \bibinfo
  {author} {\bibfnamefont {J.~X.}\ \bibnamefont {Yu}}, \bibinfo {author}
  {\bibfnamefont {P.}~\bibnamefont {Milde}}, \bibinfo {author} {\bibfnamefont
  {C.}~\bibnamefont {Huai}}, \bibinfo {author} {\bibfnamefont {H.~L.}\
  \bibnamefont {Hui}}, \bibinfo {author} {\bibfnamefont {J.~D.}\ \bibnamefont
  {Zang}}, \bibinfo {author} {\bibfnamefont {R.}~\bibnamefont {Sabirianov}},
  \bibinfo {author} {\bibfnamefont {X.~M.~M.}\ \bibnamefont {Cheng}}, \bibinfo
  {author} {\bibfnamefont {G.~X.}\ \bibnamefont {Miao}}, \bibinfo {author}
  {\bibfnamefont {H.}~\bibnamefont {Xing}}, \bibinfo {author} {\bibfnamefont
  {Y.~T.}\ \bibnamefont {Shao}}, \bibinfo {author} {\bibfnamefont {S.~A.}\
  \bibnamefont {Crooker}}, \bibinfo {author} {\bibfnamefont {L.}~\bibnamefont
  {Eng}}, \bibinfo {author} {\bibfnamefont {Y.~L.}\ \bibnamefont {Hou}},
  \bibinfo {author} {\bibfnamefont {J.~P.}\ \bibnamefont {Bird}},\ and\
  \bibinfo {author} {\bibfnamefont {H.}~\bibnamefont {Zeng}},\ }\bibfield
  {title} {\bibinfo {title} {Unconventional anomalous hall effect driven by
  self-intercalation in covalent 2d magnet cr$_2$te$_3$},\ }\href@noop {}
  {\bibfield  {journal} {\bibinfo  {journal} {Adv. Sci.}\ } (\bibinfo {year}
  {2024})}\BibitemShut {NoStop}%
\bibitem [{\citenamefont {Fujisawa}\ \emph {et~al.}(2023)\citenamefont
  {Fujisawa}, \citenamefont {Pardo-Almanza}, \citenamefont {Hsu}, \citenamefont
  {Mohamed}, \citenamefont {Yamagami}, \citenamefont {Krishnadas},
  \citenamefont {Chang}, \citenamefont {Chuang}, \citenamefont {Khoo},
  \citenamefont {Zang}, \citenamefont {Soumyanarayanan},\ and\ \citenamefont
  {Okada}}]{fujisawa2023advmater}%
  \BibitemOpen
  \bibfield  {author} {\bibinfo {author} {\bibfnamefont {Y.}~\bibnamefont
  {Fujisawa}}, \bibinfo {author} {\bibfnamefont {M.}~\bibnamefont
  {Pardo-Almanza}}, \bibinfo {author} {\bibfnamefont {C.~H.}\ \bibnamefont
  {Hsu}}, \bibinfo {author} {\bibfnamefont {A.}~\bibnamefont {Mohamed}},
  \bibinfo {author} {\bibfnamefont {K.}~\bibnamefont {Yamagami}}, \bibinfo
  {author} {\bibfnamefont {A.}~\bibnamefont {Krishnadas}}, \bibinfo {author}
  {\bibfnamefont {G.~Q.}\ \bibnamefont {Chang}}, \bibinfo {author}
  {\bibfnamefont {F.~C.}\ \bibnamefont {Chuang}}, \bibinfo {author}
  {\bibfnamefont {K.~H.}\ \bibnamefont {Khoo}}, \bibinfo {author}
  {\bibfnamefont {J.~D.}\ \bibnamefont {Zang}}, \bibinfo {author}
  {\bibfnamefont {A.}~\bibnamefont {Soumyanarayanan}},\ and\ \bibinfo {author}
  {\bibfnamefont {Y.}~\bibnamefont {Okada}},\ }\bibfield  {title} {\bibinfo
  {title} {Widely tunable berry curvature in the magnetic semimetal
  cr$_{1+\delta}$te$_2$},\ }\href@noop {} {\bibfield  {journal} {\bibinfo
  {journal} {Adv. Mater.}\ }\textbf {\bibinfo {volume} {35}} (\bibinfo {year}
  {2023})}\BibitemShut {NoStop}%
\bibitem [{\citenamefont {Song}\ \emph {et~al.}(2025)\citenamefont {Song},
  \citenamefont {Zhang}, \citenamefont {Chen}, \citenamefont {Zhang},
  \citenamefont {Cheng}, \citenamefont {Xu}, \citenamefont {Zhuang},
  \citenamefont {Sun}, \citenamefont {Zhang}, \citenamefont {Zhang},
  \citenamefont {Chen}, \citenamefont {Song}, \citenamefont {Zhang},
  \citenamefont {Zhai}, \citenamefont {Xu}, \citenamefont {Zhao}, \citenamefont
  {Zhang},\ and\ \citenamefont {Wang}}]{song2025advfuncmater}%
  \BibitemOpen
  \bibfield  {author} {\bibinfo {author} {\bibfnamefont {A.~K.}\ \bibnamefont
  {Song}}, \bibinfo {author} {\bibfnamefont {J.~E.}\ \bibnamefont {Zhang}},
  \bibinfo {author} {\bibfnamefont {Y.~Q.}\ \bibnamefont {Chen}}, \bibinfo
  {author} {\bibfnamefont {Z.~Z.}\ \bibnamefont {Zhang}}, \bibinfo {author}
  {\bibfnamefont {X.~J.}\ \bibnamefont {Cheng}}, \bibinfo {author}
  {\bibfnamefont {R.~J.}\ \bibnamefont {Xu}}, \bibinfo {author} {\bibfnamefont
  {W.~Z.}\ \bibnamefont {Zhuang}}, \bibinfo {author} {\bibfnamefont {W.~X.}\
  \bibnamefont {Sun}}, \bibinfo {author} {\bibfnamefont {Y.}~\bibnamefont
  {Zhang}}, \bibinfo {author} {\bibfnamefont {X.}~\bibnamefont {Zhang}},
  \bibinfo {author} {\bibfnamefont {Z.~Q.}\ \bibnamefont {Chen}}, \bibinfo
  {author} {\bibfnamefont {F.~Q.}\ \bibnamefont {Song}}, \bibinfo {author}
  {\bibfnamefont {Y.}~\bibnamefont {Zhang}}, \bibinfo {author} {\bibfnamefont
  {X.~C.}\ \bibnamefont {Zhai}}, \bibinfo {author} {\bibfnamefont {Y.~B.}\
  \bibnamefont {Xu}}, \bibinfo {author} {\bibfnamefont {W.~S.}\ \bibnamefont
  {Zhao}}, \bibinfo {author} {\bibfnamefont {R.}~\bibnamefont {Zhang}},\ and\
  \bibinfo {author} {\bibfnamefont {X.~F.}\ \bibnamefont {Wang}},\ }\bibfield
  {title} {\bibinfo {title} {Large anomalous hall effect in a noncoplanar
  magnetic heterostructure},\ }\href@noop {} {\bibfield  {journal} {\bibinfo
  {journal} {Adv. Funct. Mater.}\ } (\bibinfo {year} {2025})}\BibitemShut
  {NoStop}%
\bibitem [{\citenamefont {Huang}\ \emph
  {et~al.}(2021{\natexlab{b}})\citenamefont {Huang}, \citenamefont {Ma},
  \citenamefont {Wang}, \citenamefont {Li}, \citenamefont {Li}, \citenamefont
  {Xiang}, \citenamefont {Liu}, \citenamefont {Hu}, \citenamefont {Zhang},
  \citenamefont {Sun}, \citenamefont {Lu}, \citenamefont {Sheng}, \citenamefont
  {Chen}, \citenamefont {Chueh}, \citenamefont {Yang},\ and\ \citenamefont
  {Xiang}}]{huang20212dmater}%
  \BibitemOpen
  \bibfield  {author} {\bibinfo {author} {\bibfnamefont {M.}~\bibnamefont
  {Huang}}, \bibinfo {author} {\bibfnamefont {Z.~W.}\ \bibnamefont {Ma}},
  \bibinfo {author} {\bibfnamefont {S.}~\bibnamefont {Wang}}, \bibinfo {author}
  {\bibfnamefont {S.}~\bibnamefont {Li}}, \bibinfo {author} {\bibfnamefont
  {M.}~\bibnamefont {Li}}, \bibinfo {author} {\bibfnamefont {J.~X.}\
  \bibnamefont {Xiang}}, \bibinfo {author} {\bibfnamefont {P.}~\bibnamefont
  {Liu}}, \bibinfo {author} {\bibfnamefont {G.~J.}\ \bibnamefont {Hu}},
  \bibinfo {author} {\bibfnamefont {Z.~M.}\ \bibnamefont {Zhang}}, \bibinfo
  {author} {\bibfnamefont {Z.}~\bibnamefont {Sun}}, \bibinfo {author}
  {\bibfnamefont {Y.~L.}\ \bibnamefont {Lu}}, \bibinfo {author} {\bibfnamefont
  {Z.~G.}\ \bibnamefont {Sheng}}, \bibinfo {author} {\bibfnamefont
  {G.}~\bibnamefont {Chen}}, \bibinfo {author} {\bibfnamefont {Y.~L.}\
  \bibnamefont {Chueh}}, \bibinfo {author} {\bibfnamefont {S.~Y.}\ \bibnamefont
  {Yang}},\ and\ \bibinfo {author} {\bibfnamefont {B.}~\bibnamefont {Xiang}},\
  }\bibfield  {title} {\bibinfo {title} {Significant perpendicular magnetic
  anisotropy in room-temperature layered ferromagnet of cr-intercalated
  crte$_2$},\ }\href@noop {} {\bibfield  {journal} {\bibinfo  {journal} {2d
  Materials}\ }\textbf {\bibinfo {volume} {8}},\ \bibinfo {pages} {031003}
  (\bibinfo {year} {2021}{\natexlab{b}})}\BibitemShut {NoStop}%
\bibitem [{\citenamefont {Lasek}\ \emph {et~al.}(2020)\citenamefont {Lasek},
  \citenamefont {Coelho}, \citenamefont {Zberecki}, \citenamefont {Xin},
  \citenamefont {Kolekar}, \citenamefont {Li},\ and\ \citenamefont
  {Batzill}}]{lasek2020acsnano}%
  \BibitemOpen
  \bibfield  {author} {\bibinfo {author} {\bibfnamefont {K.}~\bibnamefont
  {Lasek}}, \bibinfo {author} {\bibfnamefont {P.~M.}\ \bibnamefont {Coelho}},
  \bibinfo {author} {\bibfnamefont {K.}~\bibnamefont {Zberecki}}, \bibinfo
  {author} {\bibfnamefont {Y.}~\bibnamefont {Xin}}, \bibinfo {author}
  {\bibfnamefont {S.~K.}\ \bibnamefont {Kolekar}}, \bibinfo {author}
  {\bibfnamefont {J.~F.}\ \bibnamefont {Li}},\ and\ \bibinfo {author}
  {\bibfnamefont {M.}~\bibnamefont {Batzill}},\ }\bibfield  {title} {\bibinfo
  {title} {Molecular beam epitaxy of transition metal (ti-, v-, and cr-)
  tellurides: From monolayer ditellurides to multilayer self-intercalation
  compounds},\ }\href@noop {} {\bibfield  {journal} {\bibinfo  {journal} {Acs
  Nano}\ }\textbf {\bibinfo {volume} {14}},\ \bibinfo {pages} {8473} (\bibinfo
  {year} {2020})}\BibitemShut {NoStop}%
\bibitem [{\citenamefont {Dijkstra}\ \emph {et~al.}(1989)\citenamefont
  {Dijkstra}, \citenamefont {Weitering}, \citenamefont {Vanbruggen},
  \citenamefont {Haas},\ and\ \citenamefont {Degroot}}]{dijkstra1989jpcm}%
  \BibitemOpen
  \bibfield  {author} {\bibinfo {author} {\bibfnamefont {J.}~\bibnamefont
  {Dijkstra}}, \bibinfo {author} {\bibfnamefont {H.~H.}\ \bibnamefont
  {Weitering}}, \bibinfo {author} {\bibfnamefont {C.~F.}\ \bibnamefont
  {Vanbruggen}}, \bibinfo {author} {\bibfnamefont {C.}~\bibnamefont {Haas}},\
  and\ \bibinfo {author} {\bibfnamefont {R.~A.}\ \bibnamefont {Degroot}},\
  }\bibfield  {title} {\bibinfo {title} {Band-structure calculations, and
  magnetic and transport properties of ferromagnetic chromium tellurides (crte,
  cr3te4, cr2te3)},\ }\href@noop {} {\bibfield  {journal} {\bibinfo  {journal}
  {J. Phys.: Cond. Matt.}\ }\textbf {\bibinfo {volume} {1}},\ \bibinfo {pages}
  {9141} (\bibinfo {year} {1989})}\BibitemShut {NoStop}%
\bibitem [{\citenamefont {Zhang}\ \emph {et~al.}(2023)\citenamefont {Zhang},
  \citenamefont {Liu}, \citenamefont {Zhang}, \citenamefont {Yuan},
  \citenamefont {Wen}, \citenamefont {Li}, \citenamefont {Zheng}, \citenamefont
  {Zhang}, \citenamefont {Hou}, \citenamefont {Yin}, \citenamefont {Liu},
  \citenamefont {Peng},\ and\ \citenamefont {Zhang}}]{zhang2023advmater}%
  \BibitemOpen
  \bibfield  {author} {\bibinfo {author} {\bibfnamefont {C.~H.}\ \bibnamefont
  {Zhang}}, \bibinfo {author} {\bibfnamefont {C.}~\bibnamefont {Liu}}, \bibinfo
  {author} {\bibfnamefont {J.~W.}\ \bibnamefont {Zhang}}, \bibinfo {author}
  {\bibfnamefont {Y.~Y.}\ \bibnamefont {Yuan}}, \bibinfo {author}
  {\bibfnamefont {Y.}~\bibnamefont {Wen}}, \bibinfo {author} {\bibfnamefont
  {Y.}~\bibnamefont {Li}}, \bibinfo {author} {\bibfnamefont {D.~X.}\
  \bibnamefont {Zheng}}, \bibinfo {author} {\bibfnamefont {Q.}~\bibnamefont
  {Zhang}}, \bibinfo {author} {\bibfnamefont {Z.~P.}\ \bibnamefont {Hou}},
  \bibinfo {author} {\bibfnamefont {G.}~\bibnamefont {Yin}}, \bibinfo {author}
  {\bibfnamefont {K.}~\bibnamefont {Liu}}, \bibinfo {author} {\bibfnamefont
  {Y.}~\bibnamefont {Peng}},\ and\ \bibinfo {author} {\bibfnamefont {X.~X.}\
  \bibnamefont {Zhang}},\ }\bibfield  {title} {\bibinfo {title}
  {Room-temperature magnetic skyrmions and large topological hall effect in
  chromium telluride engineered by self-intercalation},\ }\href@noop {}
  {\bibfield  {journal} {\bibinfo  {journal} {Adv. Mater.}\ }\textbf {\bibinfo
  {volume} {35}} (\bibinfo {year} {2023})}\BibitemShut {NoStop}%
\bibitem [{\citenamefont {Wang}\ \emph {et~al.}(2022)\citenamefont {Wang},
  \citenamefont {Kajihara}, \citenamefont {Matsuoka}, \citenamefont {Saika},
  \citenamefont {Yamagami}, \citenamefont {Takeda}, \citenamefont {Wadati},
  \citenamefont {Ishizaka}, \citenamefont {Iwasa},\ and\ \citenamefont
  {Nakano}}]{wang2022nanolett}%
  \BibitemOpen
  \bibfield  {author} {\bibinfo {author} {\bibfnamefont {Y.}~\bibnamefont
  {Wang}}, \bibinfo {author} {\bibfnamefont {S.}~\bibnamefont {Kajihara}},
  \bibinfo {author} {\bibfnamefont {H.}~\bibnamefont {Matsuoka}}, \bibinfo
  {author} {\bibfnamefont {B.~K.}\ \bibnamefont {Saika}}, \bibinfo {author}
  {\bibfnamefont {K.}~\bibnamefont {Yamagami}}, \bibinfo {author}
  {\bibfnamefont {Y.}~\bibnamefont {Takeda}}, \bibinfo {author} {\bibfnamefont
  {H.}~\bibnamefont {Wadati}}, \bibinfo {author} {\bibfnamefont
  {K.}~\bibnamefont {Ishizaka}}, \bibinfo {author} {\bibfnamefont
  {Y.}~\bibnamefont {Iwasa}},\ and\ \bibinfo {author} {\bibfnamefont
  {M.}~\bibnamefont {Nakano}},\ }\bibfield  {title} {\bibinfo {title}
  {Layer-number-independent two-dimensional ferromagnetism in cr3te4},\
  }\href@noop {} {\bibfield  {journal} {\bibinfo  {journal} {Nano Lett.}\
  }\textbf {\bibinfo {volume} {22}},\ \bibinfo {pages} {9964} (\bibinfo {year}
  {2022})}\BibitemShut {NoStop}%
\bibitem [{\citenamefont {Yang}\ \emph {et~al.}(2025)\citenamefont {Yang},
  \citenamefont {Ng}, \citenamefont {Zhu}, \citenamefont {Wu}, \citenamefont
  {Shi}, \citenamefont {Sun},\ and\ \citenamefont {Tang}}]{yang20252dmater}%
  \BibitemOpen
  \bibfield  {author} {\bibinfo {author} {\bibfnamefont {J.~F.}\ \bibnamefont
  {Yang}}, \bibinfo {author} {\bibfnamefont {J.~W.}\ \bibnamefont {Ng}},
  \bibinfo {author} {\bibfnamefont {C.}~\bibnamefont {Zhu}}, \bibinfo {author}
  {\bibfnamefont {Y.}~\bibnamefont {Wu}}, \bibinfo {author} {\bibfnamefont
  {J.~Y.}\ \bibnamefont {Shi}}, \bibinfo {author} {\bibfnamefont {R.~J.}\
  \bibnamefont {Sun}},\ and\ \bibinfo {author} {\bibfnamefont {B.~J.}\
  \bibnamefont {Tang}},\ }\bibfield  {title} {\bibinfo {title} {Chemical vapor
  deposition of large-area ultrathin cr3te4 nanosheets with robust
  ferromagnetism},\ }\href@noop {} {\bibfield  {journal} {\bibinfo  {journal}
  {2d Materials}\ }\textbf {\bibinfo {volume} {12}},\ \bibinfo {pages} {015002}
  (\bibinfo {year} {2025})}\BibitemShut {NoStop}%
\bibitem [{\citenamefont {Tung}\ \emph {et~al.}(2003)\citenamefont {Tung},
  \citenamefont {Kolesnichenko}, \citenamefont {Caruntu}, \citenamefont {Chou},
  \citenamefont {O'Connor},\ and\ \citenamefont {Spinu}}]{tung2003jap}%
  \BibitemOpen
  \bibfield  {author} {\bibinfo {author} {\bibfnamefont {L.~D.}\ \bibnamefont
  {Tung}}, \bibinfo {author} {\bibfnamefont {V.}~\bibnamefont {Kolesnichenko}},
  \bibinfo {author} {\bibfnamefont {D.}~\bibnamefont {Caruntu}}, \bibinfo
  {author} {\bibfnamefont {N.~H.}\ \bibnamefont {Chou}}, \bibinfo {author}
  {\bibfnamefont {C.~J.}\ \bibnamefont {O'Connor}},\ and\ \bibinfo {author}
  {\bibfnamefont {L.}~\bibnamefont {Spinu}},\ }\bibfield  {title} {\bibinfo
  {title} {Magnetic properties of ultrafine cobalt ferrite particles},\
  }\href@noop {} {\bibfield  {journal} {\bibinfo  {journal} {J. Appl. Phys.}\
  }\textbf {\bibinfo {volume} {93}},\ \bibinfo {pages} {7486} (\bibinfo {year}
  {2003})}\BibitemShut {NoStop}%
\bibitem [{\citenamefont {Mannari}(1959)}]{mannari1959ptep}%
  \BibitemOpen
  \bibfield  {author} {\bibinfo {author} {\bibfnamefont {I.}~\bibnamefont
  {Mannari}},\ }\bibfield  {title} {\bibinfo {title} {Electrical resistance of
  ferromagnetic metals},\ }\href@noop {} {\bibfield  {journal} {\bibinfo
  {journal} {Prog. Theor. Phys.}\ }\textbf {\bibinfo {volume} {22}},\ \bibinfo
  {pages} {335} (\bibinfo {year} {1959})}\BibitemShut {NoStop}%
\bibitem [{\citenamefont {Nv}\ \emph {et~al.}(1973)\citenamefont {Nv},
  \citenamefont {Dyakina},\ and\ \citenamefont
  {Startsev}}]{volkenshtein1973pssb}%
  \BibitemOpen
  \bibfield  {author} {\bibinfo {author} {\bibfnamefont {V.}~\bibnamefont
  {Nv}}, \bibinfo {author} {\bibfnamefont {V.~P.}\ \bibnamefont {Dyakina}},\
  and\ \bibinfo {author} {\bibfnamefont {V.~E.}\ \bibnamefont {Startsev}},\
  }\bibfield  {title} {\bibinfo {title} {Scattering mechanisms of conduction
  electrons in transition metals at low temperatures},\ }\href@noop {}
  {\bibfield  {journal} {\bibinfo  {journal} {Phys. Status Solidi B}\ }\textbf
  {\bibinfo {volume} {57}},\ \bibinfo {pages} {9} (\bibinfo {year}
  {1973})}\BibitemShut {NoStop}%
\bibitem [{\citenamefont {Raquet}\ \emph {et~al.}(2002)\citenamefont {Raquet},
  \citenamefont {Viret}, \citenamefont {Sondergard}, \citenamefont {Cespedes},\
  and\ \citenamefont {Mamy}}]{raquet2002prb}%
  \BibitemOpen
  \bibfield  {author} {\bibinfo {author} {\bibfnamefont {B.}~\bibnamefont
  {Raquet}}, \bibinfo {author} {\bibfnamefont {M.}~\bibnamefont {Viret}},
  \bibinfo {author} {\bibfnamefont {E.}~\bibnamefont {Sondergard}}, \bibinfo
  {author} {\bibfnamefont {O.}~\bibnamefont {Cespedes}},\ and\ \bibinfo
  {author} {\bibfnamefont {R.}~\bibnamefont {Mamy}},\ }\bibfield  {title}
  {\bibinfo {title} {Electron-magnon scattering and magnetic resistivity in
  3\textit{d} ferromagnets},\ }\href@noop {} {\bibfield  {journal} {\bibinfo
  {journal} {Phys. Rev. B}\ }\textbf {\bibinfo {volume} {66}},\ \bibinfo
  {pages} {024433} (\bibinfo {year} {2002})}\BibitemShut {NoStop}%
\bibitem [{\citenamefont {Chen}\ \emph {et~al.}(2023)\citenamefont {Chen},
  \citenamefont {Zhu}, \citenamefont {Lin}, \citenamefont {Niu}, \citenamefont
  {Liu}, \citenamefont {Zhuang}, \citenamefont {Zhang}, \citenamefont {Liang},
  \citenamefont {Sun}, \citenamefont {Chen}, \citenamefont {Hu}, \citenamefont
  {Song}, \citenamefont {Zhou}, \citenamefont {Wu}, \citenamefont {Ge},
  \citenamefont {Yang}, \citenamefont {Zhang},\ and\ \citenamefont
  {Wang}}]{chen2023advfuncmater}%
  \BibitemOpen
  \bibfield  {author} {\bibinfo {author} {\bibfnamefont {Y.~Q.}\ \bibnamefont
  {Chen}}, \bibinfo {author} {\bibfnamefont {Y.~M.}\ \bibnamefont {Zhu}},
  \bibinfo {author} {\bibfnamefont {R.~J.}\ \bibnamefont {Lin}}, \bibinfo
  {author} {\bibfnamefont {W.}~\bibnamefont {Niu}}, \bibinfo {author}
  {\bibfnamefont {R.~X.}\ \bibnamefont {Liu}}, \bibinfo {author} {\bibfnamefont
  {W.~Z.}\ \bibnamefont {Zhuang}}, \bibinfo {author} {\bibfnamefont
  {X.}~\bibnamefont {Zhang}}, \bibinfo {author} {\bibfnamefont {J.~H.}\
  \bibnamefont {Liang}}, \bibinfo {author} {\bibfnamefont {W.~X.}\ \bibnamefont
  {Sun}}, \bibinfo {author} {\bibfnamefont {Z.~Q.}\ \bibnamefont {Chen}},
  \bibinfo {author} {\bibfnamefont {Y.~S.}\ \bibnamefont {Hu}}, \bibinfo
  {author} {\bibfnamefont {F.~Q.}\ \bibnamefont {Song}}, \bibinfo {author}
  {\bibfnamefont {J.}~\bibnamefont {Zhou}}, \bibinfo {author} {\bibfnamefont
  {D.}~\bibnamefont {Wu}}, \bibinfo {author} {\bibfnamefont {B.~H.}\
  \bibnamefont {Ge}}, \bibinfo {author} {\bibfnamefont {H.~X.}\ \bibnamefont
  {Yang}}, \bibinfo {author} {\bibfnamefont {R.}~\bibnamefont {Zhang}},\ and\
  \bibinfo {author} {\bibfnamefont {X.~F.}\ \bibnamefont {Wang}},\ }\bibfield
  {title} {\bibinfo {title} {Observation of colossal topological hall effect in
  noncoplanar ferromagnet cr5te6 thin films},\ }\href@noop {} {\bibfield
  {journal} {\bibinfo  {journal} {Adv. Funct. Mater.}\ }\textbf {\bibinfo
  {volume} {33}} (\bibinfo {year} {2023})}\BibitemShut {NoStop}%
\bibitem [{\citenamefont {Zhang}\ \emph
  {et~al.}(2021{\natexlab{b}})\citenamefont {Zhang}, \citenamefont {Ambhire},
  \citenamefont {Lu}, \citenamefont {Niu}, \citenamefont {Cook}, \citenamefont
  {Jiang}, \citenamefont {Hong}, \citenamefont {Alahmed}, \citenamefont {He},
  \citenamefont {Zhang}, \citenamefont {Xu}, \citenamefont {Zhang},
  \citenamefont {Li},\ and\ \citenamefont {Bian}}]{zhang2021acsnano}%
  \BibitemOpen
  \bibfield  {author} {\bibinfo {author} {\bibfnamefont {X.~Q.}\ \bibnamefont
  {Zhang}}, \bibinfo {author} {\bibfnamefont {S.~C.}\ \bibnamefont {Ambhire}},
  \bibinfo {author} {\bibfnamefont {Q.~S.}\ \bibnamefont {Lu}}, \bibinfo
  {author} {\bibfnamefont {W.}~\bibnamefont {Niu}}, \bibinfo {author}
  {\bibfnamefont {J.}~\bibnamefont {Cook}}, \bibinfo {author} {\bibfnamefont
  {J.~S.}\ \bibnamefont {Jiang}}, \bibinfo {author} {\bibfnamefont {D.~S.}\
  \bibnamefont {Hong}}, \bibinfo {author} {\bibfnamefont {L.}~\bibnamefont
  {Alahmed}}, \bibinfo {author} {\bibfnamefont {L.}~\bibnamefont {He}},
  \bibinfo {author} {\bibfnamefont {R.}~\bibnamefont {Zhang}}, \bibinfo
  {author} {\bibfnamefont {Y.~B.}\ \bibnamefont {Xu}}, \bibinfo {author}
  {\bibfnamefont {S.~S.~L.}\ \bibnamefont {Zhang}}, \bibinfo {author}
  {\bibfnamefont {P.}~\bibnamefont {Li}},\ and\ \bibinfo {author}
  {\bibfnamefont {G.}~\bibnamefont {Bian}},\ }\bibfield  {title} {\bibinfo
  {title} {Giant topological hall effect in van der waals heterostructures of
  crte2/bi2te3},\ }\href@noop {} {\bibfield  {journal} {\bibinfo  {journal}
  {Acs Nano}\ }\textbf {\bibinfo {volume} {15}},\ \bibinfo {pages} {15710}
  (\bibinfo {year} {2021}{\natexlab{b}})}\BibitemShut {NoStop}%
\bibitem [{\citenamefont {Jeon}\ \emph {et~al.}(2022)\citenamefont {Jeon},
  \citenamefont {Na}, \citenamefont {Kim}, \citenamefont {Lee}, \citenamefont
  {Song}, \citenamefont {Kim}, \citenamefont {Park}, \citenamefont {Kim},
  \citenamefont {Noh}, \citenamefont {Kim}, \citenamefont {Jerng},\ and\
  \citenamefont {Chun}}]{jeon2022acsnano}%
  \BibitemOpen
  \bibfield  {author} {\bibinfo {author} {\bibfnamefont {J.~H.}\ \bibnamefont
  {Jeon}}, \bibinfo {author} {\bibfnamefont {H.~R.}\ \bibnamefont {Na}},
  \bibinfo {author} {\bibfnamefont {H.}~\bibnamefont {Kim}}, \bibinfo {author}
  {\bibfnamefont {S.}~\bibnamefont {Lee}}, \bibinfo {author} {\bibfnamefont
  {S.}~\bibnamefont {Song}}, \bibinfo {author} {\bibfnamefont {J.}~\bibnamefont
  {Kim}}, \bibinfo {author} {\bibfnamefont {S.}~\bibnamefont {Park}}, \bibinfo
  {author} {\bibfnamefont {J.}~\bibnamefont {Kim}}, \bibinfo {author}
  {\bibfnamefont {H.}~\bibnamefont {Noh}}, \bibinfo {author} {\bibfnamefont
  {G.}~\bibnamefont {Kim}}, \bibinfo {author} {\bibfnamefont {S.~K.}\
  \bibnamefont {Jerng}},\ and\ \bibinfo {author} {\bibfnamefont {S.~H.}\
  \bibnamefont {Chun}},\ }\bibfield  {title} {\bibinfo {title} {Emergent
  topological hall effect from exchange coupling in ferromagnetic cr2te3/
  noncoplanar antiferromagnetic cr2se3 bilayers},\ }\href@noop {} {\bibfield
  {journal} {\bibinfo  {journal} {Acs Nano}\ }\textbf {\bibinfo {volume}
  {16}},\ \bibinfo {pages} {8974} (\bibinfo {year} {2022})}\BibitemShut
  {NoStop}%
\bibitem [{\citenamefont {Nagaosa}\ \emph {et~al.}(2010)\citenamefont
  {Nagaosa}, \citenamefont {Sinova}, \citenamefont {Onoda}, \citenamefont
  {MacDonald},\ and\ \citenamefont {Ong}}]{nagaosa2010rmp}%
  \BibitemOpen
  \bibfield  {author} {\bibinfo {author} {\bibfnamefont {N.}~\bibnamefont
  {Nagaosa}}, \bibinfo {author} {\bibfnamefont {J.}~\bibnamefont {Sinova}},
  \bibinfo {author} {\bibfnamefont {S.}~\bibnamefont {Onoda}}, \bibinfo
  {author} {\bibfnamefont {A.~H.}\ \bibnamefont {MacDonald}},\ and\ \bibinfo
  {author} {\bibfnamefont {N.~P.}\ \bibnamefont {Ong}},\ }\bibfield  {title}
  {\bibinfo {title} {Anomalous hall effect},\ }\href@noop {} {\bibfield
  {journal} {\bibinfo  {journal} {Rev. Mod. Phys.}\ }\textbf {\bibinfo {volume}
  {82}},\ \bibinfo {pages} {1539} (\bibinfo {year} {2010})}\BibitemShut
  {NoStop}%
\bibitem [{\citenamefont {Tian}\ \emph {et~al.}(2009)\citenamefont {Tian},
  \citenamefont {Ye},\ and\ \citenamefont {Jin}}]{tian2009prl}%
  \BibitemOpen
  \bibfield  {author} {\bibinfo {author} {\bibfnamefont {Y.}~\bibnamefont
  {Tian}}, \bibinfo {author} {\bibfnamefont {L.}~\bibnamefont {Ye}},\ and\
  \bibinfo {author} {\bibfnamefont {X.~F.}\ \bibnamefont {Jin}},\ }\bibfield
  {title} {\bibinfo {title} {Proper scaling of the anomalous hall effect},\
  }\href@noop {} {\bibfield  {journal} {\bibinfo  {journal} {Phys. Rev. Lett.}\
  }\textbf {\bibinfo {volume} {103}},\ \bibinfo {pages} {087206} (\bibinfo
  {year} {2009})}\BibitemShut {NoStop}%
\bibitem [{\citenamefont {Hou}\ \emph {et~al.}(2015)\citenamefont {Hou},
  \citenamefont {Su}, \citenamefont {Tian}, \citenamefont {Jin}, \citenamefont
  {Yang},\ and\ \citenamefont {Niu}}]{hou2015prl}%
  \BibitemOpen
  \bibfield  {author} {\bibinfo {author} {\bibfnamefont {D.~Z.}\ \bibnamefont
  {Hou}}, \bibinfo {author} {\bibfnamefont {G.}~\bibnamefont {Su}}, \bibinfo
  {author} {\bibfnamefont {Y.}~\bibnamefont {Tian}}, \bibinfo {author}
  {\bibfnamefont {X.~F.}\ \bibnamefont {Jin}}, \bibinfo {author} {\bibfnamefont
  {S.~Y.~A.}\ \bibnamefont {Yang}},\ and\ \bibinfo {author} {\bibfnamefont
  {Q.}~\bibnamefont {Niu}},\ }\bibfield  {title} {\bibinfo {title}
  {Multivariable scaling for the anomalous hall effect},\ }\href@noop {}
  {\bibfield  {journal} {\bibinfo  {journal} {Phys. Rev. Lett.}\ }\textbf
  {\bibinfo {volume} {114}},\ \bibinfo {pages} {217203} (\bibinfo {year}
  {2015})}\BibitemShut {NoStop}%
\bibitem [{\citenamefont {Grigoryan}\ \emph {et~al.}(2017)\citenamefont
  {Grigoryan}, \citenamefont {Xiao}, \citenamefont {Wang},\ and\ \citenamefont
  {Xia}}]{grigoryan2017prb}%
  \BibitemOpen
  \bibfield  {author} {\bibinfo {author} {\bibfnamefont {V.~L.}\ \bibnamefont
  {Grigoryan}}, \bibinfo {author} {\bibfnamefont {J.}~\bibnamefont {Xiao}},
  \bibinfo {author} {\bibfnamefont {X.~H.}\ \bibnamefont {Wang}},\ and\
  \bibinfo {author} {\bibfnamefont {K.}~\bibnamefont {Xia}},\ }\bibfield
  {title} {\bibinfo {title} {Anomalous hall effect scaling in ferromagnetic
  thin films},\ }\href@noop {} {\bibfield  {journal} {\bibinfo  {journal}
  {Phys. Rev. B}\ }\textbf {\bibinfo {volume} {96}},\ \bibinfo {pages} {144426}
  (\bibinfo {year} {2017})}\BibitemShut {NoStop}%
\bibitem [{\citenamefont {Vidal}\ \emph {et~al.}(2011)\citenamefont {Vidal},
  \citenamefont {Schneider},\ and\ \citenamefont {Jakob}}]{vidal2011prb}%
  \BibitemOpen
  \bibfield  {author} {\bibinfo {author} {\bibfnamefont {E.~V.}\ \bibnamefont
  {Vidal}}, \bibinfo {author} {\bibfnamefont {H.}~\bibnamefont {Schneider}},\
  and\ \bibinfo {author} {\bibfnamefont {G.}~\bibnamefont {Jakob}},\ }\bibfield
   {title} {\bibinfo {title} {Influence of disorder on anomalous hall effect
  for heusler compounds},\ }\href@noop {} {\bibfield  {journal} {\bibinfo
  {journal} {Phys. Rev. B}\ }\textbf {\bibinfo {volume} {83}},\ \bibinfo
  {pages} {174410} (\bibinfo {year} {2011})}\BibitemShut {NoStop}%
\bibitem [{\citenamefont {Gabor}\ \emph {et~al.}(2015)\citenamefont {Gabor},
  \citenamefont {Belmeguenai}, \citenamefont {Petrisor}, \citenamefont
  {Ulhaq-Bouillet}, \citenamefont {Colis},\ and\ \citenamefont
  {Tiusan}}]{gabor2015prb}%
  \BibitemOpen
  \bibfield  {author} {\bibinfo {author} {\bibfnamefont {M.~S.}\ \bibnamefont
  {Gabor}}, \bibinfo {author} {\bibfnamefont {M.}~\bibnamefont {Belmeguenai}},
  \bibinfo {author} {\bibfnamefont {T.}~\bibnamefont {Petrisor}}, \bibinfo
  {author} {\bibfnamefont {C.}~\bibnamefont {Ulhaq-Bouillet}}, \bibinfo
  {author} {\bibfnamefont {S.}~\bibnamefont {Colis}},\ and\ \bibinfo {author}
  {\bibfnamefont {C.}~\bibnamefont {Tiusan}},\ }\bibfield  {title} {\bibinfo
  {title} {Correlations between structural, electronic transport, and magnetic
  properties of co$_2$feal$_0.5$si$_0.5$ heusler alloy epitaxial thin films},\
  }\href@noop {} {\bibfield  {journal} {\bibinfo  {journal} {Phys. Rev. B}\
  }\textbf {\bibinfo {volume} {92}},\ \bibinfo {pages} {054433} (\bibinfo
  {year} {2015})}\BibitemShut {NoStop}%
\bibitem [{\citenamefont {Meng}\ \emph {et~al.}(2017)\citenamefont {Meng},
  \citenamefont {Miao}, \citenamefont {Xu}, \citenamefont {Zhao},\ and\
  \citenamefont {Jiang}}]{meng2017pla}%
  \BibitemOpen
  \bibfield  {author} {\bibinfo {author} {\bibfnamefont {K.~K.}\ \bibnamefont
  {Meng}}, \bibinfo {author} {\bibfnamefont {J.}~\bibnamefont {Miao}}, \bibinfo
  {author} {\bibfnamefont {X.~G.}\ \bibnamefont {Xu}}, \bibinfo {author}
  {\bibfnamefont {J.~H.}\ \bibnamefont {Zhao}},\ and\ \bibinfo {author}
  {\bibfnamefont {Y.}~\bibnamefont {Jiang}},\ }\bibfield  {title} {\bibinfo
  {title} {Thickness dependence of magnetic anisotropy and intrinsic anomalous
  hall effect in epitaxial co$_2$mnal film},\ }\href@noop {} {\bibfield
  {journal} {\bibinfo  {journal} {Phys. Lett. A}\ }\textbf {\bibinfo {volume}
  {381}},\ \bibinfo {pages} {1202} (\bibinfo {year} {2017})}\BibitemShut
  {NoStop}%
\bibitem [{\citenamefont {Kikkawa}\ \emph {et~al.}(2015)\citenamefont
  {Kikkawa}, \citenamefont {Uchida}, \citenamefont {Daimon}, \citenamefont
  {Qiu}, \citenamefont {Shiomi},\ and\ \citenamefont
  {Saitoh}}]{kikkawa2015prb}%
  \BibitemOpen
  \bibfield  {author} {\bibinfo {author} {\bibfnamefont {T.}~\bibnamefont
  {Kikkawa}}, \bibinfo {author} {\bibfnamefont {K.}~\bibnamefont {Uchida}},
  \bibinfo {author} {\bibfnamefont {S.}~\bibnamefont {Daimon}}, \bibinfo
  {author} {\bibfnamefont {Z.~Y.}\ \bibnamefont {Qiu}}, \bibinfo {author}
  {\bibfnamefont {Y.}~\bibnamefont {Shiomi}},\ and\ \bibinfo {author}
  {\bibfnamefont {E.}~\bibnamefont {Saitoh}},\ }\bibfield  {title} {\bibinfo
  {title} {Critical suppression of spin seebeck effect by magnetic fields},\
  }\href@noop {} {\bibfield  {journal} {\bibinfo  {journal} {Phys. Rev. B}\
  }\textbf {\bibinfo {volume} {92}},\ \bibinfo {pages} {064413} (\bibinfo
  {year} {2015})}\BibitemShut {NoStop}%
\bibitem [{\citenamefont {Irkhin}\ and\ \citenamefont
  {Irkhin}(2021)}]{irkhin2021condmat}%
  \BibitemOpen
  \bibfield  {author} {\bibinfo {author} {\bibfnamefont {V.~Y.}\ \bibnamefont
  {Irkhin}}\ and\ \bibinfo {author} {\bibfnamefont {Y.~P.}\ \bibnamefont
  {Irkhin}},\ }\href@noop {} {\bibinfo {title} {Electronic structure,
  correlation effects and physical properties of d- and f-metals and their
  compounds}} (\bibinfo {year} {2021}),\ \Eprint
  {https://arxiv.org/abs/cond-mat/9812072} {arXiv:cond-mat/9812072 [cond-mat]}
  \BibitemShut {NoStop}%
\bibitem [{\citenamefont {Doi}\ \emph {et~al.}(1970)\citenamefont {Doi},
  \citenamefont {Nakao},\ and\ \citenamefont {Kamimura}}]{doi1970jpsj}%
  \BibitemOpen
  \bibfield  {author} {\bibinfo {author} {\bibfnamefont {T.}~\bibnamefont
  {Doi}}, \bibinfo {author} {\bibfnamefont {K.}~\bibnamefont {Nakao}},\ and\
  \bibinfo {author} {\bibfnamefont {H.}~\bibnamefont {Kamimura}},\ }\bibfield
  {title} {\bibinfo {title} {Valence band structure of tellurium .1. k.p
  perturbation method},\ }\href@noop {} {\bibfield  {journal} {\bibinfo
  {journal} {J. Phys. Soc. Jpn.}\ }\textbf {\bibinfo {volume} {28}},\ \bibinfo
  {pages} {36} (\bibinfo {year} {1970})}\BibitemShut {NoStop}%
\bibitem [{\citenamefont {Furukawa}\ \emph {et~al.}(2017)\citenamefont
  {Furukawa}, \citenamefont {Shimokawa}, \citenamefont {Kobayashi},\ and\
  \citenamefont {Itou}}]{furukawa2017ncomm}%
  \BibitemOpen
  \bibfield  {author} {\bibinfo {author} {\bibfnamefont {T.}~\bibnamefont
  {Furukawa}}, \bibinfo {author} {\bibfnamefont {Y.}~\bibnamefont {Shimokawa}},
  \bibinfo {author} {\bibfnamefont {K.}~\bibnamefont {Kobayashi}},\ and\
  \bibinfo {author} {\bibfnamefont {T.}~\bibnamefont {Itou}},\ }\bibfield
  {title} {\bibinfo {title} {Observation of current-induced bulk magnetization
  in elemental tellurium},\ }\href@noop {} {\bibfield  {journal} {\bibinfo
  {journal} {Nat. Commun.}\ }\textbf {\bibinfo {volume} {8}},\ \bibinfo {pages}
  {954} (\bibinfo {year} {2017})}\BibitemShut {NoStop}%
\bibitem [{\citenamefont {Sakano}\ \emph {et~al.}(2020)\citenamefont {Sakano},
  \citenamefont {Hirayama}, \citenamefont {Takahashi}, \citenamefont {Akebi},
  \citenamefont {Nakayama}, \citenamefont {Kuroda}, \citenamefont {Taguchi},
  \citenamefont {Yoshikawa}, \citenamefont {Miyamoto}, \citenamefont {Okuda},
  \citenamefont {Ono}, \citenamefont {Kumigashira}, \citenamefont {Ideue},
  \citenamefont {Iwasa}, \citenamefont {Mitsuishi}, \citenamefont {Ishizaka},
  \citenamefont {Shin}, \citenamefont {Miyake}, \citenamefont {Murakami},
  \citenamefont {Sasagawa},\ and\ \citenamefont {Kondo}}]{sakano2020prl}%
  \BibitemOpen
  \bibfield  {author} {\bibinfo {author} {\bibfnamefont {M.}~\bibnamefont
  {Sakano}}, \bibinfo {author} {\bibfnamefont {M.}~\bibnamefont {Hirayama}},
  \bibinfo {author} {\bibfnamefont {T.}~\bibnamefont {Takahashi}}, \bibinfo
  {author} {\bibfnamefont {S.}~\bibnamefont {Akebi}}, \bibinfo {author}
  {\bibfnamefont {M.}~\bibnamefont {Nakayama}}, \bibinfo {author}
  {\bibfnamefont {K.}~\bibnamefont {Kuroda}}, \bibinfo {author} {\bibfnamefont
  {K.}~\bibnamefont {Taguchi}}, \bibinfo {author} {\bibfnamefont
  {T.}~\bibnamefont {Yoshikawa}}, \bibinfo {author} {\bibfnamefont
  {K.}~\bibnamefont {Miyamoto}}, \bibinfo {author} {\bibfnamefont
  {T.}~\bibnamefont {Okuda}}, \bibinfo {author} {\bibfnamefont
  {K.}~\bibnamefont {Ono}}, \bibinfo {author} {\bibfnamefont {H.}~\bibnamefont
  {Kumigashira}}, \bibinfo {author} {\bibfnamefont {T.}~\bibnamefont {Ideue}},
  \bibinfo {author} {\bibfnamefont {Y.}~\bibnamefont {Iwasa}}, \bibinfo
  {author} {\bibfnamefont {N.}~\bibnamefont {Mitsuishi}}, \bibinfo {author}
  {\bibfnamefont {K.}~\bibnamefont {Ishizaka}}, \bibinfo {author}
  {\bibfnamefont {S.}~\bibnamefont {Shin}}, \bibinfo {author} {\bibfnamefont
  {T.}~\bibnamefont {Miyake}}, \bibinfo {author} {\bibfnamefont
  {S.}~\bibnamefont {Murakami}}, \bibinfo {author} {\bibfnamefont
  {T.}~\bibnamefont {Sasagawa}},\ and\ \bibinfo {author} {\bibfnamefont
  {T.}~\bibnamefont {Kondo}},\ }\bibfield  {title} {\bibinfo {title} {Radial
  spin texture in elemental tellurium with chiral crystal structure},\
  }\href@noop {} {\bibfield  {journal} {\bibinfo  {journal} {Phys. Rev. Lett.}\
  }\textbf {\bibinfo {volume} {124}},\ \bibinfo {pages} {136404} (\bibinfo
  {year} {2020})}\BibitemShut {NoStop}%
\bibitem [{\citenamefont {Holstein}\ and\ \citenamefont
  {Primakoff}(1940)}]{holstein1940pr}%
  \BibitemOpen
  \bibfield  {author} {\bibinfo {author} {\bibfnamefont {T.}~\bibnamefont
  {Holstein}}\ and\ \bibinfo {author} {\bibfnamefont {H.}~\bibnamefont
  {Primakoff}},\ }\bibfield  {title} {\bibinfo {title} {Field dependence of the
  intrinsic domain magnetization of a ferromagnet},\ }\href@noop {} {\bibfield
  {journal} {\bibinfo  {journal} {Phys. Rev.}\ }\textbf {\bibinfo {volume}
  {58}},\ \bibinfo {pages} {1098} (\bibinfo {year} {1940})}\BibitemShut
  {NoStop}%
\bibitem [{\citenamefont {Kuz'min}\ \emph {et~al.}(2020)\citenamefont
  {Kuz'min}, \citenamefont {Skokov}, \citenamefont {Diop}, \citenamefont
  {Radulov},\ and\ \citenamefont {Gutfleisch}}]{kuzmin2020epjp}%
  \BibitemOpen
  \bibfield  {author} {\bibinfo {author} {\bibfnamefont {M.~D.}\ \bibnamefont
  {Kuz'min}}, \bibinfo {author} {\bibfnamefont {K.~P.}\ \bibnamefont {Skokov}},
  \bibinfo {author} {\bibfnamefont {L.~V.~B.}\ \bibnamefont {Diop}}, \bibinfo
  {author} {\bibfnamefont {I.~A.}\ \bibnamefont {Radulov}},\ and\ \bibinfo
  {author} {\bibfnamefont {O.}~\bibnamefont {Gutfleisch}},\ }\bibfield  {title}
  {\bibinfo {title} {Exchange stiffness of ferromagnets},\ }\href@noop {}
  {\bibfield  {journal} {\bibinfo  {journal} {Eur. Phys. J. Plus}\ }\textbf
  {\bibinfo {volume} {135}},\ \bibinfo {pages} {301} (\bibinfo {year}
  {2020})}\BibitemShut {NoStop}%
\end{thebibliography}%

\end{document}